%% file: ghost.tex
\documentclass[smallextended]{svjour3}
\usepackage{mathptmx}
\usepackage{amsmath}
\usepackage{color}
\usepackage{graphicx}
\usepackage{subcaption}
\usepackage{listings}
\usepackage{courier}
\usepackage{epstopdf}
\usepackage{picins}
\usepackage{todonotes}
\usepackage{url}
\usepackage{sidecap}
\usepackage{bm}
\usepackage{algpseudocode}
\usepackage{tabto}
\usepackage{hyperref}

\makeatletter
\def\cl@chapter{\@elt {theorem}}
\makeatother

\usepackage[capitalise]{cleveref}

\usepackage{enumerate}
\graphicspath{{./img/}}
\input{GHmacros}

\definecolor{lightgray}{gray}{0.95}
\colorlet{lstcomment}{purple!40!black}
\lstdefinestyle{MyCStyle}{language=C,
    frame=tb,
    backgroundcolor=\color{lightgray},
    basicstyle=\scriptsize\ttfamily,
    keywordstyle=\bfseries\color{green!40!black},
    commentstyle=\itshape\color{lstcomment},
    identifierstyle=\color{blue},
    stringstyle=\color{orange},
    morekeywords={ghost_lidx,ghost_gidx,ghost_task,ghost_task_flags,ghost_task_cur,ghost_task_create,ghost_task_enqueue,ghost_task_wait,ghost_task_destroy,ghost_spmv,__m256d,_mm256_store_pd,_mm256_load_pd,_mm256_set1_pd,_mm256_setzero_pd,_mm256_mul_pd},
    escapeinside={<@}{@>}
}
\lstdefinestyle{MyBashStyle}{language=bash,
    frame=tb,
    backgroundcolor=\color{lightgray},
    basicstyle=\scriptsize\ttfamily,
    escapeinside={<@}{@>}}
\newcommand{\nvidia}{NVIDIA}
\newcommand{\sellcs}{SELL-C-$\sigma$}

\newcommand{\new}[1]{#1}%\color{blue}{#1} \color{black}}
\newcommand{\ghost}{\mbox{GHOST}}

\begin{document}
\title{\ghost: Building Blocks for High Performance\\ Sparse Linear Algebra on
Heterogeneous Systems}
\titlerunning{\ghost: High Performance Sparse Linear Algebra on Heterogeneous
Systems}
%\thanks{This work was supported by the
%German Research Foundation (DFG) through the Priority Programs 1648
%``Software for Exascale Computing'' (SPPEXA) under project ESSEX (``Equipping Sparse
%Solvers for Exascale'').}}
%\author{Moritz Kreutzer\footnotemark[2]\footnotetext[2]{Erlangen Regional Computing Center,
%Friedrich-Alexander-Universit\"at Erlangen-N\"urnberg, 91058 Erlangen,
%Germany}\and Jonas Thies\footnotemark[3]\footnotetext[3]{German Aerospace Center (DLR), Simulation and Software Technology, 51147 K\"oln, Germany} \and
%    Melven R\"ohrig-Z\"ollner\footnotemark[3]\and Andreas
%    Pieper\footnotemark[4]\footnotetext[4]{Institute of Physics,
%    Ernst-Moritz-Arndt-Universit\"at Greifswald, 17489 Greifswald, Germany} \and
%    Faisal Shahzad\footnotemark[2] \and Martin
%    Galgon\footnotemark[5]\footnotetext[5]{Bergische Universit\"at Wuppertal,
%    42097 Wuppertal, Germany} \and Achim
%Basermann\footnotemark[3]\and Holger Fehske\footnotemark[4]\and Georg Hager\footnotemark[2]\and Gerhard
%Wellein\footnotemark[2]}
\author{Moritz Kreutzer \and Jonas Thies \and Melven R\"ohrig-Z\"ollner \and
Andreas Pieper \and Faisal Shahzad \and Martin Galgon \and Achim Basermann \and
Holger Fehske \and Georg Hager \and Gerhard Wellein}

\institute{
    M. Kreutzer \and F. Shahzad \and G. Hager \at
Erlangen Regional Computing Center, Friedrich-Alexander-Universit\"at
Erlangen-N\"urnberg, 91058 Erlangen,
Germany\\\email{\{moritz.kreutzer, faisal.shahzad, georg.hager\}@fau.de} \and J. Thies \and M. R\"ohrig-Z\"ollner
\and A. Basermann \at German Aerospace Center (DLR), Simulation and Software
Technology, 51147 K\"oln, Germany\\\email{\{jonas.thies,
melven.roehrig-zoellner, achim.basermann\}@dlr.de} \and A. Pieper \and H. Fehske \at Institute of Physics,
Ernst-Moritz-Arndt-Universit\"at Greifswald, 17489 Greifswald,
Germany\\\email{\{pieper, fehske\}@physik.uni-greifswald.de} \and
Martin Galgon \at Bergische Universit\"at Wuppertal, 42097 Wuppertal,
Germany\\\email{galgon@math.uni-wuppertal.de}
\and G. Wellein \at Department of Computer Science, Friedrich-Alexander-Universit\"at
Erlangen-N\"urnberg, 91058 Erlangen, Germany\\\email{gerhard.wellein@fau.de} }

\maketitle

%%%%%%%%%%%%%%%%%%%%%%%%%%%%%%%%%%%%%%%%%%%%%%%%%%%%%%%%%%%%%%%%%%%%%%%%%
\begin{abstract}
While many of the architectural details of future exascale-class high
performance computer systems are still a matter of intense research,
there appears to be a general consensus that they will be strongly
heterogeneous, featuring ``standard'' as well as ``accelerated''
resources. Today, such resources are available as multicore
processors, graphics processing units (GPUs), and other accelerators such as the Intel Xeon
Phi. Any software infrastructure that claims usefulness for such
environments must be able to meet their inherent challenges: massive
multi-level parallelism, topology, asynchronicity, and
abstraction. The ``General, Hybrid, and Optimized Sparse Toolkit'' (\ghost)
is a collection of building blocks that targets algorithms dealing with sparse matrix
representations on current and future large-scale systems. 
It implements the ``MPI+X'' paradigm, has a pure C interface, and provides
hybrid-parallel numerical kernels,
intelligent resource management, and truly heterogeneous parallelism for multicore CPUs, Nvidia GPUs,
and the Intel Xeon Phi. We describe the details of its design with respect to the
challenges posed by modern heterogeneous
supercomputers and recent algorithmic developments. 
Implementation details which
are indispensable for achieving high efficiency are pointed out and their
necessity is justified by performance
measurements or predictions based on performance models.
The library code and several applications are available as open source. 
\new{We also provide instructions on how to make use of \ghost\ in
existing software packages, together with a case study which
demonstrates the applicability and performance of \ghost\ as a component within a larger
software stack.}
\end{abstract}
%%%%%%%%%%%%%%%%%%%%%%%%%%%%%%%%%%%%%%%%%%%%%%%%%%%%%%%%%%%%%%%%%%%%%%%%%

\section{Introduction and related work}

\subsection{Sparse solvers on heterogeneous hardware}
\label{sec:intro}
\input{ghost_intro.tex}

\subsection{Related work}
\label{sec:related_work}
\input{ghost_related_work.tex}

\subsection{Contribution}
\label{sec:contribution}
\input{ghost_contribution.tex}

\subsection{Testbed}
\label{sec:testbed}
\input{ghost_testbed.tex}

\section{Design principles}
\label{sec:design_principles}
\input{ghost_design_principles.tex}

\subsection{Supported architectures and programming models}
\label{sec:architectures}
\input{ghost_architectures.tex}

\subsection{Parallelism in \ghost}
\label{sec:data_parallelism}
\input{ghost_data_parallelism.tex}

\section{Available data structures}
\label{sec:data_structures}
\input{ghost_data_structures.tex}

\subsection{Sparse matrices}
\label{sec:sparsemat}
\input{ghost_sparsemat.tex}

\subsection{Dense matrices}
\label{sec:densemat}
\input{ghost_densemat.tex}

\section{Runtime features}
\label{sec:runtime_features}
\input{ghost_runtime_features.tex}

\subsection{Transparent and data-parallel heterogeneous execution}
\label{sec:heterogeneous_execution}
\input{ghost_heterogeneous_execution.tex}

\subsection{Affinity-aware resource management}
\label{sec:resource_management}
\input{ghost_resource_management.tex}

\section{Performance features}
\label{sec:performance_features}
\input{ghost_performance_features.tex}

\subsection{Sparse matrix storage}
\label{sec:sparsemat_storage}
\input{ghost_sparsemat_storage.tex}

\subsection{Block vectors}
\label{sec:blockvectors}
\input{ghost_blockvectors.tex}

\subsection{Kernel fusion}
\label{sec:kernelfusion}
\input{ghost_kernelfusion.tex}

\subsection{Low-level implementation and code generation}
\label{sec:implementation}
\input{ghost_implementation.tex}

\section{Using \ghost\ in existing iterative solver packages}
\label{sec:ghost_integration}
\input{ghost_integration.tex}

\subsection{\new{Case study: An eigensolver with Trilinos and \ghost}}
\label{sec:case_study}
\input{ghost_case_study.tex}

%\section{Case study: A fault-tolerant CG solver for multiple right hand sides}
%\todo{Do we have the space and need for a case study? Also: Does a case study
%make sense and generate more insight?}
%\label{sec:ghost_case_study}
%\input{ghost_case_study.tex}
  
\section{Conclusion and outlook}

\subsection{Conclusions}
\label{sec:conclusions}
\input{ghost_conclusions.tex}

\subsection{Outlook}
\label{sec:outlook}
\input{ghost_outlook.tex}

\begin{acknowledgements}
This work was supported by the
German Research Foundation (DFG) through the Priority Program 1648
``Software for Exascale Computing'' (SPPEXA) under project ESSEX (``Equipping Sparse
Solvers for Exascale'').
We would like to thank Intel Germany and Nvidia for providing test systems for
benchmarking. Special thanks go to Andreas Alvermann for providing sparse matrix
generation functions for testing and everyone else who contributed to \ghost{},
directly or indirectly.
\end{acknowledgements}

\bibliographystyle{spmpsci}
\bibliography{ghost}
\end{document}

%% file: GHmacros.tex
\newcommand{\bq}{\begin{equation}}
\newcommand{\eq}{\end{equation}}

\newcommand{\doi}[1]{\href{http://dx.doi.org/#1}{DOI:#1}}

\newcommand{\construction}[1]{}%\marginpar{\includegraphics*[width=1cm]{Under_construction_icon-blue.eps}{#1}}}

%% file: ghost_intro.tex
Users of modern supercomputers are facing several obstacles on their way to highly 
efficient software. Out of those, probably the most prominent is the ever increasing
level of parallelism in hardware architectures. Increasing the parallelism on
the chips -- both in terms of the number of cores as well as inside the
core itself -- is currently the only way to increase the maximum performance while keeping the
energy consumption at a reasonable level. 
The parallelization of hardware architectures peaks in the use of
accelerators, coprocessors, 
or graphics
processing units (GPUs) for general purpose computations. Those devices trade
off core sophistication against a very high core count, achieving an extremely high level of parallelism with unmatched
peak floating point performance per Watt.
Today, more than 20\% of all TOP500~\cite{TOP500} systems are heterogeneous and the accelerators in
those installations account for almost a third of the entire aggregated
TOP500 performance. This evolution has led to the
emergence of a large
scientific community dealing with various aspects of accelerator programming and
a considerable amount of accelerator-enabled software packages.
However, ``heterogeneous software'' often means
``accelerator software''. A fact which is frequently not being considered is that also the CPU part of a heterogeneous
system can contribute significantly to a program's performance.
On the other hand, even in simpler CPU-only machines, properties like memory 
hierarchy, ccNUMA effects and
thread affinity play an important role in high performance code development.
The addition of accelerators further increases the level of complexity in the system
topology. At the same time, the world's largest compute 
clusters exhibit a steady growth in terms of cores and nodes as a result of
continuously increasing requirements from application users. This poses
challenges to algorithms and software in terms of scalability. 

There is a wide range of applications which require computation with very
large sparse matrices. The development of sparse linear and eigenvalue solvers that achieve extreme parallelism 
therefore remains an important field of research. 
One example is the analysis of novel materials like graphene~\cite{Pieper14} and
topological insulators~\cite{Schubert11} in the field of solid state physics, which is a driving
force in the \ghost\ development within the ESSEX project\footnote{\url{http://http://blogs.fau.de/essex}}.
\new{These eigenvalue problems require the computational power of full petascale systems for many hours, so it is crucial
to achieve optimal performance at all levels.}

Recent work in the area of sparse matrix algorithms
can roughly be subdivided into three categories. Methods that increase computational 
intensity (i.e. reduce/avoid communication), methods that hide communication by overlapping 
it with computation, and fully asynchronous algorithms. To the first category
belong, among others,
block Krylov methods and the communication avoiding GMRES 
(CA-GMRES~\cite{hoemmenIPDPS08}) method, which require optimized block vector 
kernels. The second category includes the pipelined CG and GMRES 
methods~\cite{ghysels2012hiding}. An example for the third category 
is the asynchronous ILU preconditioner by Chow and Patel~\cite{chow15fine_grained}. Methods 
from the latter two categories benefit from an easy-to-use tasking model that 
delivers high performance. \new{The novel implementation of ILU methods in~\cite{chow15fine_grained} replaces
the poorly scaling forward/backward substitution by a matrix polynomial, increasing the performance requirements
of the sparse matrix-vector multiplication in preconditioned Krylov methods.}

%% file: ghost_related_work.tex
%\begin{itemize}
%    \item MAGMA
%    \item StarPU
%    \item Paralution
%    \item Task vs. data parallelism in heterogeneous environments
%    \item BLAS 2.5
%    \item ViennaCL
%\end{itemize}
There is a large interest in efficient heterogeneous software driven by
the developments in modern supercomputer architectures described above. Many
efforts follow a task-parallel approach, which strives to map a heterogeneous 
workload to heterogeneous hardware in an efficient way. 
The most prominent software package
implementing task-parallel heterogeneous execution is MAGMA~\cite{MAGMA}. A
major drawback of MAGMA is the absence of built-in MPI support, i.e., users have to implement MPI parallelism around
MAGMA on their own.
Under the hood, MAGMA uses the StarPU runtime system as
proposed by Augnoett et al.~\cite{Augonnet11} for 
automatic task-based work scheduling on heterogeneous systems. 
%Belviranli et
%al.~\cite{Belviranli13} proposed a dynamic scheduling and workload balancing scheme called
%HDSS (Heterogeneous Dynamic Self Scheduler) which is similar
%to StarPU.
%\todo{We probably don't need the Belvirani cite}
Another significant attempt towards heterogeneous software is ViennaCL~\cite{Rupp10}.
Being based on CUDA, OpenCL or OpenMP, this software package can execute the same code on a wide
range of compute architectures. However, concurrent use of different
architectures \new{for a single operation} is not supported. Besides, ViennaCL has limited support for complex
numbers, which is problematic for many applications. The same applies
to the C++ framework LAMA~\cite{LAMA}, a library based on MPI+OpenMP/CUDA with special focus on
large sparse matrices. \new{PETSc~\cite{PETSc} is an MPI-parallel library for
the scalable solution of scientific applications modeled by partial
differential equations. Its intended programming model is pure MPI, with MPI+X
support for GPUs (`X' being CUDA or OpenCL) and some limited support for
threading. It also lacks support of heterogeneous computation of single
operations.} 
Another library containing sparse iterative solvers and preconditioners is
PARALUTION~\cite{PARALUTION}. However, the multi-node and complex number support
is restricted to the non-free version of this software.
\new{The Trilinos packages Kokkos and Tpetra~\cite{baker2012tpetra} implement an MPI+X 
approach similar to the one used in \ghost. 
Being implemented in C++, they clearly separate the MPI level (Tpetra package) from the node level (Kokkos package), 
whereas \ghost\ can benefit from tighter integration for, e.g., improved asynchronous MPI communication 
(cf. \cref{sec:resource_management}).
In \cref{sec:case_study} we will provide a performance comparison of \ghost\
vs. Trilinos for an eigenvalue solver.

}

%The main differences are:
%\begin{enumerate}[(i)]
%\item Trilinos uses recent C++ 11 features, whereas the \ghost~interface is pure
%    C which allows straightforward interoperability with Fortran codes.
    %\todo{Is this a requirement for Fortran interoperability?}
%\item Trilinos clearly separates the MPI level (Tpetra package) from the node level (Kokkos package), 
%whereas \ghost\ uses its tasking mechanism for improved asynchronous MPI communication 
%(cf.~Section~\ref{sec:resource_management}).
%\item Trilinos has a focus on generality over optimal performance. For instance, the central 
%sparse matrix format is CRS, whereas \ghost~uses the SELL-C-$\sigma$ format for good 
%performance on both CPU and GPU (cf. Section \ref{sec:sparsemat_storage}). 
%On the other 
%hand, the Kokkos programming model allows users to write hybrid parallel functionality 
%themselves.
%, whereas \ghost\ does not strive to be as easily extensible.
%This is similar to user-defined tasks in \ghost\, while the Kokkos approach appears to be
%simpler from an implementation point of view.
%\todo{Formulation...}
%\end{enumerate}
%\todo{obigen Text weniger negativ formulieren}
\new{While all of these libraries certainly improve the accessibility of heterogeneous hardware to a wide
range of applications, they do not fit our purpose of extreme scale eigenvalue computations in an optimal
way. In particular, we believe that a single library for building blocks integrating well-tuned kernels, 
communication on all levels and good performance on heterogeneous systems `out of the box' is key to 
satisfying the needs of scientists who are trying to tackle problems at the edge of `what can be done'.}

%There are several publications about heterogeneous execution of various
%algorithms, including SpMV, using multiple GPUs in a single node. For example, Verschoor et
%al.~\cite{Verschoor12} implement the Conjugate Gradient method (CG) and
%Yamazaki et al.~\cite{Yamazaki14} present a task-parallel heterogeneous CPU/multi-GPU
%implementation of the CA-GMRES algorithm. They utilize multiple GPUs within a single compute
%node for the matrix powers and orthogonalization kernels and eventually solve the least squares problem 
%on CPUs.
%\todo{Georg: "Hier geht es ja um konzeptionelle Ähnlichkeiten, also würde ich die
%    Algorithmen uU gar nicht erwähnen. Außerdem haben wir ja hier gar keine
%    volle Applikation, da wollen wir nicht drauf rumreiten, dass andere das
%haben."
%Andererseits gibt es im Repository ja einige Algorithmen zum Download...}

%Yang et al.~\cite{Yang11} have presented an MPI-parallel multi-GPU SpMV which
%follows a data-parallel approach similar to the strategy used in GHOST. However, they mention the
%limited scalability of their approach due to redundant communication of vector
%data. Additionally, they do not include CPUs in their computation.
%This is done by Cardellini et al.~\cite{Cardellini14} in their heterogeneous, data-parallel and MPI-enabled implementation of SpMV.

%% file: ghost_contribution.tex
% several approaches to heterogeneous execution driven by modern architectures
% mainly task-parallel
% GHOST is the first approach to a truly data-parallel heteregenous sparse solver toolkit
% unifies low-level optimization driven by performance engineering into a rather high-level framework
In this work, we present the software package \ghost\ (\emph{General, Hybrid and Optimized Sparse
Toolkit}).
As summarized in \cref{sec:related_work}, there is a range of efforts towards
efficient sparse linear algebra on heterogeneous hardware driven by modern
hardware architectural developments. 
\ghost\ can be classified as an approach towards a highly scalable, and
truly heterogeneous sparse linear algebra toolkit
%which features data parallelism between heterogeneous hardware architectures 
with a key target in the development process being
optimal performance on all parts of heterogeneous systems.
In close collaboration with experts from the application side we focus on a few key operations often needed in sparse eigenvalue solvers and
provide highly optimized and performance-engineered implementations for those.
We show that disruptive changes of data structures may be necessary to achieve
efficiency on modern CPUs and accelerators featuring wide single instruction
multiple data (SIMD) units and
multiple cores.

\new{
%    One may wonder if it is worth the effort to optimize a few core operations for the last 10 or 20\%
%of performance. 
%We argue that on multi-million dollar systems, computing time is a valuable resource
%and those operations that are carried out over and over in an algorithm should be as fast as possible.
%We focus on these high end systems in the Petaflop range and beyond, but of course the kernels we provide
%can be used on smaller clusters or single workstations as well.
One may argue about whether performance should be the primary goal in a CS\&E
software library and whether it is worth the effort to optimize a few core
operations for a two-digit percentage gain in performance.
Our efforts are targeted at large-scale supercomputers in the petaflop range and
beyond, and computing time is a valuable resource there. Even a performance gain
below an order of magnitude can become significant in terms of time, energy, and
money spent on the large scale. 
Needless to say, the kernels we provide can be used on smaller clusters or single
workstations as well.
\ghost\ does not give up generality or extensibility for this purpose, rather we
aim to provide performance-optimized (guided by performance models)
kernels for some commonly used algorithms (e.~g. the kernel polynomial
method~\cite{Kreutzer15}, the block Jacobi-Davidson method~\cite{Roehrig14} or
Chebyshev filter diagonalization~\cite{Pieper15}).
The successful implementations of these methods (which are very popular in fields like material physics and
quantum chemistry) will serve as blueprints for other techniques such as advanced preconditioners needed in other
CS\&E disciplines. In the application areas we consider right now, methods such as incomplete factorization or multigrid
can usually not be applied straightforwardly. The matrices that appear may not have an interpretation as physical quantities
discretized on a mesh, they may be completely indefinite, and they may have relatively small diagonal entries and/or
random elements~\cite{Galgon2015}.}

%We demonstrate that library developers often have to abandon goals like generality and
%consistency of data structures if the main objective is high efficiency.
%\todo{Umformulieren. ``Generality'' haben wir natuerlich, steht ja auch im
%Namen. Bzgl. Datenstrukturen: Manchmal muss man ``disruptive'' Aenderungen der
%Codebasis vornehmen um gute Performance zu erreichen.
%Insbesondere zaehlt das fuer Datenstrukturen: CRS -> SELLCS, col-major -> rowmajor
%etc.}
%\todo{Abschw\"achen wg. Kokkos/Tpetra, s.o.. MK: "Jonas, ist das Problem von
%Kokkos nur schlechte Performance? Kann man dort auch eine SpMV datenparallel mit
%CPUs und GPUs loesen?"}

A key feature of \ghost\ is the transparency to the user when it comes to
heterogeneous execution. In contrast to other heterogeneous software packages
(cf. \cref{sec:related_work}), \ghost\ uses a data-parallel
approach for work distribution among compute devices. While a task-parallel
approach is well-suited for workloads with complex dependency graphs, the data-parallel approach used by \ghost\ may
be favorable for uniform workloads (i.e., algorithms where all parts have
similar resource requirements) or algorithms where an efficient task-parallel
implementation is unfeasible. On the process level, \ghost's tasking mechanism
still allows for flexible work distribution beyond pure data parallelism. 

\ghost\ unifies optimized low-level kernels whose development process is being
guided by performance modelling into a high-level toolkit which allows
resource-efficient execution on modern heterogeneous hardware. 
Note that for uniform workloads, performance models
like the roofline model~\cite{Williams09} are a suitable tool to check an
implementation's efficiency.
In recent work,
\ghost\ has proven to scale up to the petaflop level, extending the
scaling studies presented in~\cite{Kreutzer15}. 
%Special attention
%is being paid to the efficient utilization of all parts of a heterogeneous system.
%\ghost\ should be considered a toolbox containing building blocks for
%large-scale sparse linear algebra problems. 
A list of challenges we are
addressing specifically and the corresponding sections in this paper can be given as follows:
\begin{enumerate}[(i)]
    \item Emerging asynchronous sparse solver algorithms are in need of a
        light-weight, affinity-aware and threading-friendly task-based execution
        model. In this context, the high relevance of OpenMP should be noted,
        which requires the tasking model to be compatible to OpenMP-parallel
        codes. See \cref{sec:resource_management}.
%        \todo{Talk about the relevance of OpenMP and why we need to support it
%        before?}
    \item Existing software rarely uses all components of heterogeneous systems in an
        efficient manner. See \cref{sec:heterogeneous_execution,sec:sparsemat_storage}.
    \item The potential performance of compute libraries is often limited by the
        requirement of high generality, leading to a lack of
        application-specific kernels. See \cref{sec:kernelfusion}.
    \item The possibilities for application developers to feed their knowledge into
        compute libraries for higher performance are often limited. See \cref{sec:implementation}.
    \item The applicability of optimization techniques like vector blocking is
        often limited due to restrictions in existing data structures. Fundamental
        changes to data structures are often hard to integrate in existing
        software packages. See \cref{sec:blockvectors,sec:sparsemat_storage}.
\end{enumerate}

\ghost\ is available as a BSD-licensed open source download~\cite{GHOST}.
Along with it, a list of sample application based on \ghost\ (e.g., a Conjugate
Gradient solver and a Lanczos eigensolver) can be downloaded.
On top of that, the iterative solver package and sister project of \ghost\ named PHIST~\cite{PHIST} 
can use \ghost\ to execute more sophisticated algorithms like, e.g., the block
Jacobi-Davidson eigensolver as described in~\cite{Roehrig14}, and blocked
versions of the MinRes and GMRES linear solvers.

%% file: ghost_testbed.tex
All experiments in this paper have been conducted on the Emmy\footnote{\url{http://www.rrze.fau.de/dienste/arbeiten-rechnen/hpc/systeme/emmy-cluster.shtml}} cluster located
at the Erlangen Regional Computing
Center.
\Cref{tab:architectures} summarizes the hardware components used in this
cluster.
The Intel C/C++ compiler in version 14 and CUDA in version 6.5 have been used
for compilation. Intel MKL 11.1 was used as the BLAS library on the CPU.

\begin{table}[tb]
\centering
\begin{tabular}{llccccc}
\hline
\textbf{Alias} & \textbf{Model} & \textbf{Clock} & \textbf{SIMD} &
\textbf{Cores/} & $\bm{b}$ & $\bm{P^{\mathrm{peak}}}$ \\
               &       & \textbf{(MHz)} & \textbf{(Bytes)} & \textbf{SMX} & \textbf{(GB/s)} & \textbf{(Gflop/s)} \\
\hline
\hline
CPU & Intel Xeon E5-2660 v2 & 2200 & 32  & 10 & 50  & 176 \\
GPU & Nvidia Tesla K20m     & 706  & $128\dots512^*$ & 13 & 150 & 1174 \\ 
PHI & Intel Xeon Phi 5110P  & 1050 & 64  & 60 & 150 & 1008 \\
\hline
\end{tabular}
\caption{Relevant properties of all architectures used in this paper. The
    attainable memory bandwidth as measured with the STREAM~\cite{McCalpin95}
    benchmark is denoted by $b$ and $P^{\mathrm{peak}}$ is the theoretical
    peak floating point performance. Turbo mode
was activated on the CPU and the GPU was configured with ECC enabled.\\
$^*$: SIMD processing is done by 32 threads. Hence, the SIMD width in bytes depends on
the data type in use: 128 bytes is valid for 4-byte (single precision floating
point) data while 512 bytes
corresponds to complex double precision data.}
\label{tab:architectures}
\end{table}
%\todo{Explain b and ppeak in caption}

%% file: ghost_design_principles.tex
In this section, fundamental design decisions of the \ghost\ development are
discussed and justified. This includes the support of certain hardware
architectures as well as fundamental parallelization paradigms.

%% file: ghost_architectures.tex
Many modern compute clusters comprise heterogeneous nodes.
Usually, such nodes consist of one or more multi-core CPUs and one or more
accelerator cards. In the TOP500~\cite{TOP500} systems of November 2014, 96\% of all
accelerator performance stems from \nvidia\ Fermi/Kepler GPUs or Intel Xeon Phi
(to be called ``PHI'') coprocessors. Hence, we decided to limit the
accelerator support in GHOST to those two architectures. Instead of implementing
support for any kind of hardware architecture, our primary goal is to stick to
the dominant platforms and to develop properly performance-engineered code for
those.

Although GPUs and PHIs share the name \emph{accelerators}, there are significant
differences in how those devices are operated.
GPUs can only be driven in \emph{accelerator mode}, i.e., data transfers and 
compute kernels must be launched explicitly from a main program running on a host CPU.
The PHI can be operated in accelerator mode, too. However, in addition to
this, the PHI can also be driven in \emph{native mode}, i.e.,
in the same way as a multicore CPU would be used. 
In \ghost\, only native
execution on the PHI is supported, i.e., the PHI hosts its own process. 
Hence, the PHI can be considered as a CPU node on its own.
With regard to the PHI as
a multi-(many-)core CPU, it has to be taken into
account that serial code may run at very low performance due
to its very simple core architecture. 

%In the rest of this paper, the term CPU
%will be used even when talking about the Intel Xeon Phi.
%\todo{Is this true? We use the term PHI afterwards as well}

%% file: ghost_data_parallelism.tex
\begin{figure}
\centering
\begin{subfigure}[b]{.40\textwidth}
    \includegraphics[width=\textwidth]{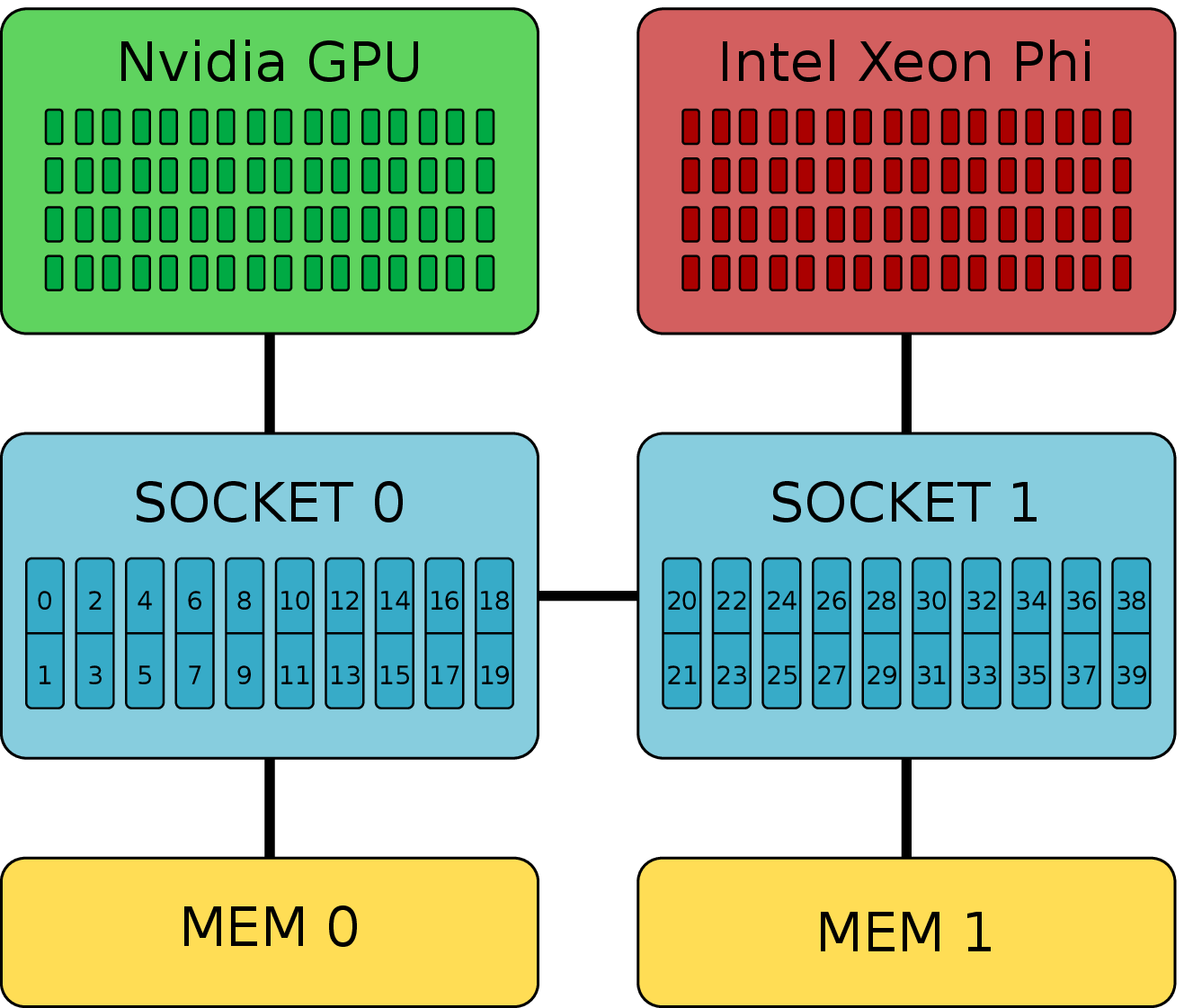}
    \caption{Heterogeneous node}
    \label{fig:heterogeneous_node}
\end{subfigure}
\hfill
\begin{subfigure}[b]{.40\textwidth}
    \includegraphics[width=\textwidth]{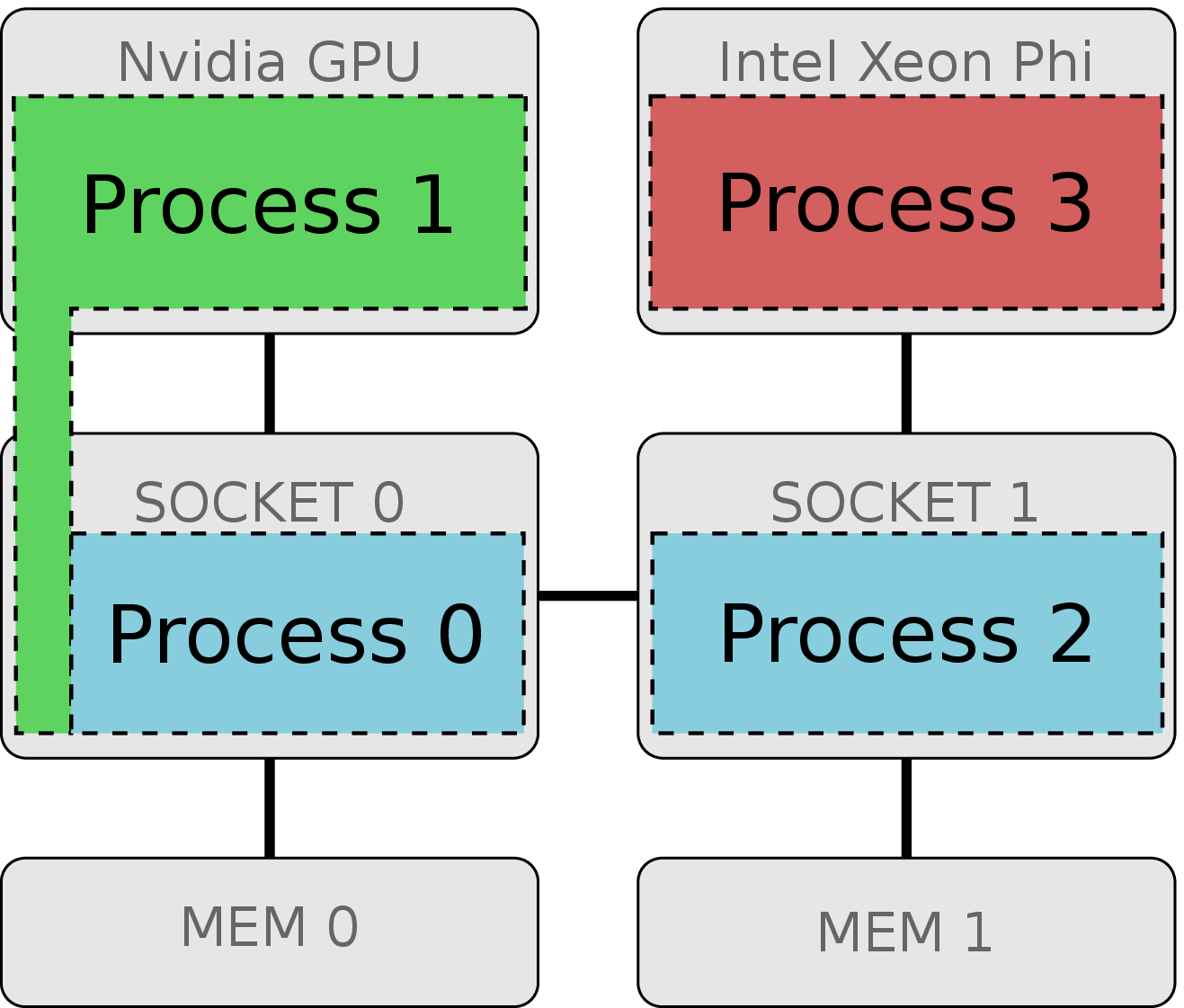}
    \caption{Process placement}
    \label{fig:heterogeneous_node_procs}
\end{subfigure}
\caption{Heterogeneous compute node and sensible process placement as suggested
by \ghost.}
\end{figure}
For illustration of the principles, we consider a heterogeneous node as shown in
\cref{fig:heterogeneous_node}. This node contains two multicore CPU
sockets with ten cores and two-way hyper-threading each. In total,
there are 20 hardware threads or processing units (PUs) available per socket. In
addition to that, one GPU and one PHI are
attached to the node as accelerators. Note that a node with two different accelerator
architectures is unlikely to be installed in a production system. 
%However,
%a handful of such nodes is installed in the Emmy cluster.

In terms of parallelization, \ghost\ implements the ``MPI+X''
paradigm, i.e., coarse-grained parallelism is done by means of MPI
accompanied with fine-grained and device-specific parallelization
mechanisms (``X''). One may certainly omit the ``X'' part and go with
plain MPI altogether if the hardware can be efficiently utilized in
this way; the plain fact that modern hardware exhibits complex
topologies does not mean that a hybrid programming model is required
in all cases, and it may sometimes even be
counterproductive.
However, interesting opportunities in terms of load
balancing, additional levels of parallelism, communication hiding,
etc., arise from combining MPI with a threading
model~\cite{10.1109/PDP.2009.43}. This is why
\ghost\ supports OpenMP for the ``X'' component on CPUs. Further down
the hardware hierarchy, implicit vectorization by compiler-friendly
code and pragmas as well as explicit vectorization using Single
Instruction Multiple Data (SIMD) intrinsics provide efficient
single-threaded code. On Nvidia GPUs, CUDA is used as the
``X'' parallelization layer.

In general, parallelism in compute applications can be categorized into \emph{data}
and \emph{task} parallelism. The term \emph{data parallelism} describes a number of workers
operating on the
same data set, each having assigned a certain amount of work.
The term \emph{task parallelism} describes workers working on independent tasks at the same time.
\ghost\ implements both data (between processes) as described in
\cref{sec:heterogeneous_execution} and task parallelism (inside a
process) as analyzed in \cref{sec:resource_management}.

In many cases, algorithms from sparse linear algebra are centered around a
single and potentially large sparse
system matrix. Hence, the distribution of work in \ghost\ is done in a
matrix-centered way. More precisely, the system matrix is distributed row-wise across
the MPI processes.
The amount of work per process can either be expressed by the number of rows
or the number of nonzero elements. 
Details on the implementation are given in
\cref{sec:heterogeneous_execution}.

%% file: ghost_data_structures.tex
There are two major data structures in \ghost, namely sparse and dense matrices
(\texttt{ghost\_sparsemat} and \texttt{ghost\_densemat}, respectively). 
Dense vectors are represented as dense matrices with a single column. 
\new{Both data structures implement a row-wise distribution among MPI processes.
We do not support 2D partitionings of these data structures or direct conversion
routines between them, but this may be added in the future. 
}

%% file: ghost_sparsemat.tex
\ghost\ supports the \sellcs\ sparse matrix storage format as introduced in 
\cite{Kreutzer15}. 
\new{Note that this is not necessarily a restriction, as the well-known CRS 
storage format can be expressed as SELL-1-1. Further special cases of \sellcs\ will be
listed in \cref{sec:sparsemat_storage}.} 
More details on sparse matrix storage are given in
\cref{sec:sparsemat_storage}.
As mentioned in \cref{sec:data_parallelism}, the sparse system matrix is 
the central data structure in \ghost.

A significant performance bottleneck for highly scalable sparse solvers may be the
generation of the system matrix. In \ghost\, this matrix can be stored in a
file, either in the Matrix Market \cite{MMFormat} format or a binary format
which resembles the CRS data format. 
%Reading in a Matrix Market file cannot be
%done in a scalable way because the row order is not defined for this file
%format. 
%For the binary format, however, a single process reads the row length information from
%the file and the row share for each process is determined. Afterwards, each
%process reads only the portion of the file which belongs to its matrix share.
%Note that if the measure of work is the number of rows (instead
%of the number of nonzero elements), the (serial) read-in
%of row length information can be omitted.
\new{However, the scalability of this approach is intrinsically
limited. The preferred method of matrix construction in \ghost\ is via a
callback function provided by the user, which allows to construct a matrix row
by row.}
%There is another and potentially even more scalable way for matrix creation in
%\ghost{}. The user defines a callback function from which \ghost\ creates the
%matrix row by row.
The function must have the following signature:
\begin{lstlisting}[style=MyCStyle]
int mat(ghost_gidx row, ghost_lidx *len, ghost_gidx *col, void *val, void *arg);
\end{lstlisting}
\ghost{} passes the global matrix row to the function. The user should then
store the number of non-zeros of this row in the argument \texttt{rowlen} and
the column indices and values of the non-zeros in \texttt{col} and
\texttt{val}. \new{Any further arguments can be passed in \texttt{arg}.}
The maximum number of non-zeros must be set in advance such that
\ghost\ can reserve enough space for the \texttt{col} and \texttt{val} arrays.
%\todo{Is the stuff about sparse matrix generation too technical?}

There are several reasons which make it necessary to permute rows of the sparse
system matrix.
A global (inter-process) permutation of matrix rows can be applied in order to
minimize the communication volume of, e.g., the sparse matrix vector
multiplication (SpMV) kernel and to enforce more
cache-friendly memory access patterns. Currently, \ghost\
can be linked against PT-SCOTCH~\cite{Chevalier08} for this purpose. A matrix' row lengths
and column indices are passed to PT-SCOTCH which results in a
permutation vector on each process containing global indices. Afterwards, the
matrix is assembled on each process according to the global permutation.
Our experiments revealed that this approach is limited in terms of scalability.
For that reason, we are going to include support for more global permutation schemes 
that improve communication reduction
in future work, such as the parallel hypergraph partitioner as implemented in
Zoltan~\cite{Devine06}.

In addition to the global permutation, a local (intra-process) permutation can
be applied, e.g., to minimize
the storage overhead of the \sellcs\ sparse matrix format (cf.
\cref{sec:sparsemat_storage}. 
Another
potential reason for a local matrix permutation is row coloring. \ghost\ has the possibility
to permute a sparse matrix according to a coloring scheme obtained from
ColPack~\cite{Gebremedhin13}. This kind of re-ordering may be
necessary for the parallelization of, e.g., the Kaczmarz~\cite{Kaczmarz37}
algorithm or a Gau{\ss}-Seidel smoother as present in the HPCG benchmark.
%\todo{also necessary for matrix-based Jacobi as it is used in HPCG (?)}

\new{Note that an application-based permutation, e.g., by optimizing the numbering of
nodes in a mesh-based problem, usually leads to better overall performance and should be preferred over an a posteriori
permutation, e.g., with PT-SCOTCH. In \ghost\, the former can be achieved by the
user by a sensible implementation of the matrix construction via the callback
interface.}

%% file: ghost_densemat.tex
\ghost{} is a framework for sparse linear algebra. Dense matrices are mainly
occurring as dense vectors (dense matrices with a single column) or blocks of
dense vectors (to be referred to as \emph{block vectors}).
\Cref{sec:blockvectors} will cover the aspect of block vectors in more detail. Block vectors can be considered 
as tall and skinny dense matrices,
i.e., dense matrices with a large row count (in the dimension of the sparse
system matrix) but relatively few (at most a few hundred) columns.
\new{Furthermore, a \texttt{ghost\_densemat} can be used to represent small local or replicated 
matrices, e.g. the result of an inner product of two (distributed) block vectors.}
\piccaption{Views of a dense matrix.\label{fig:densemat_views}}
\parpic[r]{\includegraphics[width=.45\textwidth]{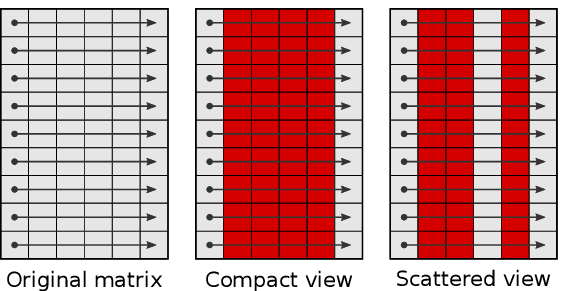}}
%\begin{wrapfigure}{R}{0.4\textwidth}
%\centering
%\begin{subfigure}[b]{.12\textwidth}
%    \includegraphics[width=\textwidth]{densemat_source.eps}
%    \caption{Original matrix}
%    \label{fig:densemat_source}
%\end{subfigure}
%\hfill
%\begin{subfigure}[b]{.12\textwidth}
%    \includegraphics[width=\textwidth]{densemat_view_compact.eps}
%    \caption{Compact view}
%    \label{fig:densemat_view_compact}
%\end{subfigure}
%\hfill
%\begin{subfigure}[b]{.12\textwidth}
%    \includegraphics[width=\textwidth]{densemat_view_scattered.eps}
%    \caption{Scattered view}
%    \label{fig:densemat_view_scattered}
%\end{subfigure}
%\caption{Views of dense matrices}
%\label{fig:densemat_views}
%\end{wrapfigure}
Instead of allocating its own memory, a dense matrix can also be created as a \emph{view} of another
dense matrix or a view of arbitrary data in memory.
This makes it easily possible to let a function work on a sub-matrix or a subset
of vectors in a larger block vector without having to copy any data.
Additionally, by viewing ``raw'' data in memory it is possible to integrate \ghost{} into existing code 
(cf.~\cref{sec:ghost_integration}). 
A potential disadvantage of using non-\ghost{} data structures is the 
violation of data alignment which may result in a performance loss.
\ghost{} implements different kinds of views, as shown in \cref{fig:densemat_views}. In
general, compact views allow vectorized computation with the matrix
data. This is not the case for scattered views due to the ``gaps'' in memory layout in the
leading dimension caused by columns not included in the view. In this case, it may be favorable to create a compact clone
of the scattered view before executing the computation.

Dense matrices can be chosen to be stored in a (locally) row- or column-major
manner. In many cases, row-major storage (which corresponds to interleaved
storage of block vectors) yields better performance and should be preferred over
column-major storage (cf. \cref{sec:blockvectors}). On the other hand, column-major storage may be required
for easy integration of \ghost{} into existing software. \ghost{} offers mechanisms to change 
the storage layout either in-place or out-of-place, while copying a block vector.

%% file: ghost_runtime_features.tex
In this section we describe runtime features which are deeply woven into the
software architecture and constitute \ghost's unique feature set. 
In contrast to the so-called \emph{performance} features which will be
introduced in \cref{sec:performance_features}, they are fundamentally
built into the library and hard to apply to other approaches.

%% file: ghost_heterogeneous_execution.tex
The distribution of work among the heterogeneous components is done on a per-process basis 
where each process (MPI rank) is bound to a fixed set of PUs.
This allows flexible scaling and adaption to various kinds of heterogeneous
systems.
The sets of PUs on a single node are disjoint, i.e., the compute resources are
exclusively available to a process.
For the example node shown in \cref{fig:heterogeneous_node}, the minimum amount of processes for
heterogeneous execution on the full node is three. 

Application developers are frequently confused by the ccNUMA memory
structure of modern compute nodes and how to handle it to avoid
performance penalties. Although the required programming strategies
are textbook knowledge today,
establishing perfect local memory
access may be tricky if a multithreaded process spans multiple
ccNUMA domains even if proper thread-core affinity is in place
and parallel first-touch initialization is performed~\cite{Hager:2010:IHP:1855048}.
A simple way to avoid ccNUMA
problems is to create one process per multicore CPU socket, which would result
in a process count of four as illustrated for the example node in
\cref{fig:heterogeneous_node_procs}.
Processes 0 and 2 cover one CPU socket each.  
Process 1 drives the GPU. As this has to be done in accelerator mode, this process also occupies one core of the host system. 
Note that this core is located on the socket whose PCI Express bus the GPU is
attached to and thus has to be subtracted from Process 0's CPU set.
Process 3 is used for the PHI. The process can directly be located on the
accelerator which is used in native mode, i.e., no host resources are used for
driving the PHI.

For each numerical function in the \ghost\ library, implementations for the different architectures are present.
However, the choice of the specific
implementation of a kernel does not have to be made by the user. Consequently, in almost all usage cases, no changes to the code
are necessary when switching between different hardware architectures. An
exception of this rule is, e.g., the creation of a dense matrix view from plain
data: If the
dense matrix is located on a GPU, the plain data must be valid GPU
memory.

An intrinsic property of heterogeneous systems is that the components differ in
terms of performance. For efficient heterogeneous execution it is
important that the performance differences get reflected in the work
distribution. In \ghost\, the underlying sparse system matrix gets divided on a
row-wise basis among
all processes. 
For example, if component A is
expected to have twice the performance of component B, process A will get assigned a
twice as large chunk (either in terms of rows or in terms of nonzero elements) of the system matrix as process B.
\begin{figure}
\includegraphics[width=\columnwidth]{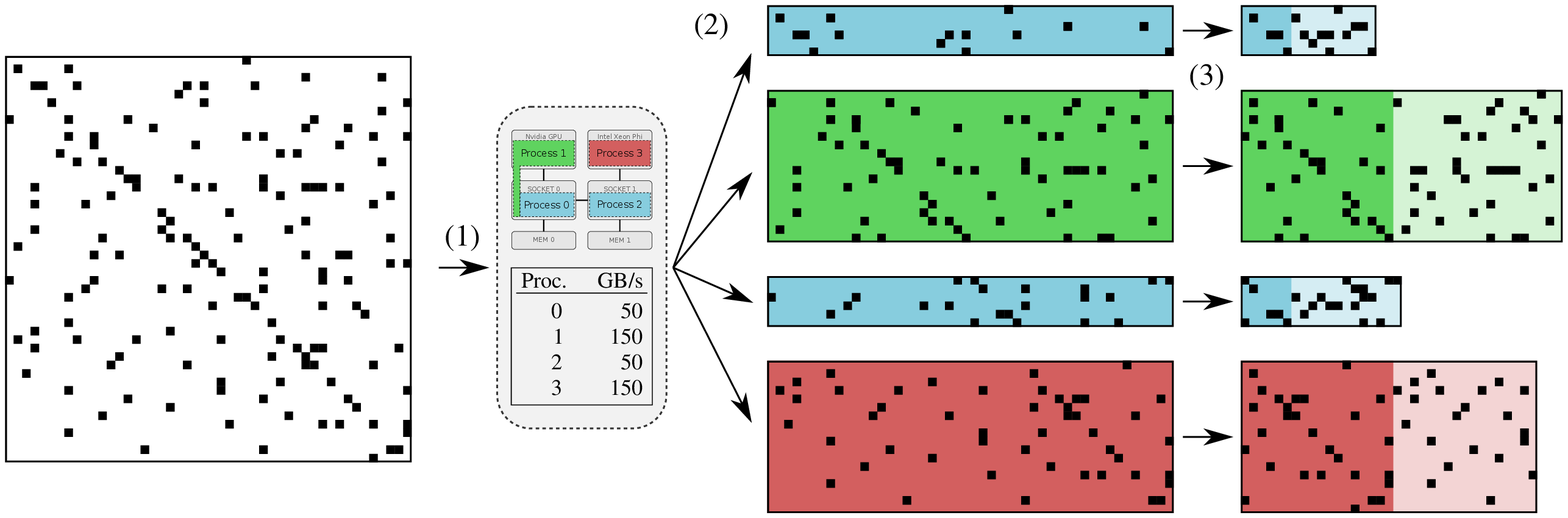}
\caption{Heterogeneous row-wise distribution of a sparse matrix. Step (1) is the
determination of process weights according to the device's peak memory
bandwidths. In step (2), a partial sparse matrix is created on each process. In
order to avoid integer overflows and storage overhead in the communication
buffers, 
the column indices of elements in the remote matrix part (pale colors) are compressed in step (3).}
\label{fig:sparsemat_procs}
\end{figure}
\Cref{fig:sparsemat_procs} illustrates the row-wise distribution of a sparse
%\todo{Ist der Begriff ``halo buffer'' bekannt?}
matrix among the example processes shown in
\cref{fig:heterogeneous_node_procs}. 
As the performance of sparse solvers is often bound by main
memory bandwidth, the device-specific maximum attainable bandwidth, as
given in \cref{tab:architectures}, has been chosen as the work distribution
criterion in this example. Note that an arbitrary work share for each
process/architecture can easily be specified at runtime.

Internally, each process gets assigned a \emph{type} which allows to define the compute platform used by an executable.
Valid types are \texttt{CPU} and \texttt{GPU}.
The type can be set explicitly at runtime either via API calls or by specifying an environment variable.
If multiple processes are launched on a node containing CPUs and GPUs, the type gets selected automatically if not explicitly specified.
In this case, Process 0 is always of type \texttt{CPU}, initially covering all CPUs in the node.
Processes $1$ to $N$ are of type
\texttt{GPU} where $N$ is equal to the number of GPUs attached to the node.
For each GPU process getting added to a node, a small CPU set (usually a
single core) gets subtracted from Process 0's resources.
If any more than (1 + ``Number of GPUs'') processes get placed on a node, the addition of any further processes causes a division of Process 0's CPU set into equally sized smaller CPU sets.
A good number of processes to put on a node is (``Number of CPUs'' + ``Number of
GPUs''), which is an easy way to avoid NUMA locality problems by having one
process per CPU socket.

In the following we demonstrate the heterogeneous execution capabilities on our
example node using a simple program which measures the SpMV performance for a given matrix
and storage format (downloadable from the \ghost\
website~\cite{GHOST}). In this case, we used the \texttt{Janna/ML\_Geer}
matrix
\footnote{\url{http://www.cise.ufl.edu/research/sparse/matrices/Janna/ML_Geer.html}}
(dimension $n=1,504,002$, number of non-zeros $n_\mathrm{nz} = 110,686,677$)
stored in SELL-32-1. 
\new{Performance will be reported in Gflop/s, with 1 Gflop/s 
corresponding to a minimum memory bandwidth of 6 GByte/s. 
This relation is founded on the
minimum code balance of the SpMV kernel.}
If we want to perform computations on the CPU only and use one process per CPU
socket, the
type has to be set explicitly and a suitable number of processes has to be
launched on the host: 
\begin{lstlisting}[style=MyBashStyle]
> <@\textbf{GHOST\_TYPE=CPU}@> mpiexec <@\textbf{-nopin -np 2 -ppn 2}@> ./spmvbench -v -m ML_Geer.mtx \
  -f SELL-32-1
<@\textcolor{red}{[GHOST] PERFWARNING: The number of MPI ranks (1) on this node is not optimal!}@>
                     <@\textcolor{red}{Suggested number: 3 (2 NUMA domains + 1 CUDA device)}@>
<@\textcolor{red}{[GHOST] PERFWARNING: There is 1 Xeon Phi in the set of active nodes but only 0}@>
                     <@\textcolor{red}{are used!}@>
Region      | Calls |    P_max | P_skip10
-----------------------------------------
spmv (GF/s) |   100 | 1.64e+01 | 1.64e+01
\end{lstlisting}
The overall number of processes is set via the \texttt{-np} flag and the number
of processes per host is set using \texttt{-ppn}.
Note that automatic thread/process pinning by the MPI startup script has been
suppressed by \texttt{-nopin}.
This should always be done to avoid conflicts with
\ghost's resource management.
%Launching the executable with a single process and not specifying further options we obtain the following output:
%\begin{lstlisting}[style=MyBashStyle]
%> ./spmvbench -v -m ML_Geer.mtx -f SELL-32-1
%...
%...
%Region      | Calls |    P_max | P_skip10
%-----------------------------------------
%spmv (GF/s) |   100 | 1.38e+01 | 1.35e+01
%\end{lstlisting}
The maximum performance over all 100 runs is given in \texttt{P\_max}.
\texttt{P\_skip10} shows the average performance over all but the first ten
iterations.
The performance warnings (omitted in the following listings) issued by \ghost\ indicate that the node is not used to its full
extent. The suggested process count of three is in accordance with the knowledge about
the node architecture; each node contains two CPU sockets and one GPU.
The Intel PHI attached to this node has to be considered as a node on its
own.
%The warnings will not be printed in the following examples.
%Additionally, as mentioned above, using a single process on a NUMA
%system like this may result in some performance penalty.
%\todo{Georg: "Dieser genze Abschnitt fällt etwas mit der Tür ins Haus. Klar kann
%    NUMA ein Problem sein, ist es aber inm Beispiel weiter unten offenbar nicht.
%    Auch ist nicht klar, wie die -np Option von mpiexec mit dem GHOST resource
%mgmt interagiert."}
The achieved performance of 16.4 Gflop/s matches the prediction of a simple roofline 
model for this algorithm and two CPU sockets.
If the example program should use the GPU for computation, the following command
has to be invoked:

\noindent\begin{minipage}{\textwidth} %prevent line break
\begin{lstlisting}[style=MyBashStyle]
> <@\textbf{GHOST\_TYPE=GPU}@> ./spmvbench -v -m ML_Geer.mtx -f SELL-32-1
Region      | Calls |    P_max | P_skip10
-----------------------------------------
spmv (GF/s) |   100 | 2.28e+01 | 2.27e+01
\end{lstlisting}
\end{minipage}
From the single-device runs we can easily deduce that the
GPU execution was 2.75 times as fast as the execution on a single CPU socket.
In the current implementation the weight, i.e., the amount of work, assigned to
each process in heterogeneous runs, has to
be specified manually. Future work will include automatic weight selection based
on micro-benchmarks and dynamic adaption of weights at runtime (cf.~\ref{sec:outlook}). 
Starting the example program using three processes and a work ratio between CPU
and GPU of 1:2.75 yields the following:
\begin{lstlisting}[style=MyBashStyle]
> mpiexec -nopin <@\textbf{-np 3 -ppn 3}@> ./spmvbench -v -m ML_Geer.mtx -f SELL-32-1 \
  <@\textbf{-w 1:2.75}@>
<@\textcolor{blue}{[GHOST] PE0 INFO: Setting GHOST type to CPU.}@>
<@\textcolor{blue}{[GHOST] PE1 INFO: Setting GHOST type to GPU.}@>
<@\textcolor{blue}{[GHOST] PE2 INFO: Setting GHOST type to CPU.}@>
Region      | Calls |    P_max | P_skip10
-----------------------------------------
spmv (GF/s) |   100 | 3.11e+01 | 3.09e+01
\end{lstlisting}
The information log messages indicate that the process types have automatically been set as described above.
The achieved performance is less than the accumulated single-device
performances. This is due to the MPI communication of input vector data which is done in each SpMV
iteration.
For testing purposes, it is possible to suppress the communication by selecting
an appropriate ``pseudo SpMV''
routine. Note that this does not give the correct result for the SpMV
operation if the input vector data changes between successive iterations.
\begin{lstlisting}[style=MyBashStyle]
> mpiexec -nopin -np 3 -ppn 3 ./spmvbench -v -m ML_Geer.mtx -f SELL-32-1 \
  -w 1:2.75 <@\textbf{-s nocomm}@>
...
Region      | Calls |    P_max | P_skip10
-----------------------------------------
spmv (GF/s) |   100 | 3.85e+01 | 3.73e+01
\end{lstlisting}
Now, the heterogeneous performance sums up to the accumulated single-device
performances.
In order to include the node's PHI in the computation, the library and executable have to be compiled for
the MIC architecture, resulting in an additional executable file
\texttt{./spmvbench.mic}. For setting up heterogeneous execution using the PHI, the following has to be done:
\begin{lstlisting}[style=MyBashStyle]
> # Assemble machine file for three MPI ranks on the host and one on the PHI
> echo -e "$(hostname -s):3\n$(hostname -s)-mic0:1" > machinefile
> export I_MPI_MIC=1             # Enable MPI on the PHI
> export I_MPI_MIC_POSTFIX=.mic  # Specify the postfix of the MIC executable
\end{lstlisting}
Using all parts of the heterogeneous node for computing the ``pseudo SpMV'',
the following is obtained:
\begin{lstlisting}[style=MyBashStyle]
> mpiexec -nopin <@\textbf{-np 4 -machinefile machinefile}@> \
  ./spmvbench -v -m ML_Geer.mtx -f SELL-32-1 <@\textbf{-w 1:2.75:2.75 -s nocomm}@>
...
Region      | Calls |    P_max | P_skip10
-----------------------------------------
spmv (GF/s) |   100 | 5.64e+01 | 5.47e+01
\end{lstlisting}
%\todo{Should we include a PHI only run here?}
The PHI got assigned the same share of work as the GPU. Note that there may
still be optimization potential regarding the load balance.
The total node performance adds up to approximately 55 Gflop/s, which indicates
a good use of the aggregated memory bandwidth of all resources.
If the actual SpMV function is computed, the optimal weights are slightly
different, due to a higher communication effort on the GPU and PHI.
\begin{lstlisting}[style=MyBashStyle]
> mpiexec -nopin -np 4 -machinefile machinefile \
  ./spmvbench -v -m ML_Geer.mtx -f SELL-32-1 <@\textbf{-w 1:2.3:2.1}@>
...
Region      | Calls |    P_max | P_skip10
-----------------------------------------
spmv (GF/s) |   100 | 3.97e+01 | 3.88e+01
\end{lstlisting}
Note that the inclusion of the PHI barely leads to a performance benefit
over the\linebreak CPU+GPU run.
This is due to the small amount of work on each device and an increasing
dominance of communication over the slow PCI express bus as a result of this.
The impact of communication may be reduced by matrix re-ordering (cf.
\cref{sec:sparsemat}) or more sophisticated communication mechanisms (using
GPUdirect, pipelined communication, etc., cf. \cref{sec:outlook}).

%% file: ghost_resource_management.tex
Although \ghost\ follows a data-parallel approach across processes for heterogeneous execution,
work is organized in tasks on the process level. The increasing
asynchronicity of algorithms together with the necessity for sensible hardware affinity and the avoidance of
resource conflicts constitute the need for
a unit of work which is
aware of the underlying hardware has to be used: a \ghost\ task. Affinity and
resource management is implemented by means of the \texttt{hwloc}
library~\cite{Broquedis10} which is, besides a BLAS library, the only build
dependency of \ghost.

There are existing solutions for task parallelism. Apart from the ones named in
\cref{sec:related_work}, OpenMP tasks or Intel's Threading Building Blocks can
be mentioned here. However, a crucial requirement in the design phase of \ghost\
was to support affinity-aware OpenMP parallelism in user-defined tasks. As we
could not find a light-weight existing solution which meets our requirements, we decided to implement an
appropriate tasking model
from scratch. For example, both Intel TBB and Cilk Plus warn about using those frameworks together with OpenMP in their user manuals. 
This is due to potential performance issues caused by core over-subscription.
As OpenMP is in widespread use in scientific codes, this limitation
disqualifies the integration of TBB and Cilk Plus tasks for many existing applications.
Note that \ghost\ tasks lack a list of features compared to existing solutions,
such as intelligent resolution of dependencies in complex scenarios. Yet, for
most of our usage scenarios they work well enough.
\new{In our opinion, a holistic performance engineering approach is a key to
optimal performance for complex scenarios. Thus, we decided to make the resource
management a part of the \ghost\ library. 
%If future circumstances
%require a separation of this component from \ghost, one could consider to soften
%the holism if it
%does not compromise performance and usability.
}

%\todo{A small word about possible issues with tasking and OpenMP}

Generally speaking, a task's work can be an
arbitrary user-defined function in which OpenMP can be used without having
to worry about thread affinity or resource conflicts. The threads of a task are pinned to exclusive
compute resources, if not specified otherwise. 
\ghost\ tasks are used in the library itself, e.g., for explicitly overlapping
communication and computation in a parallel SpMV (see below). However, the
mechanism is designed in a way that allows easy integration of the tasking
capabilities also into user code.
The user-relevant properties of a task are as follows:
\begin{lstlisting}[style=MyCStyle]
typedef struct {
    void * (*func) (void *); /* callback function where work is done */
    void * arg; /* arguments to the work function */
    void * ret; /* return value of the work function */
    struct ghost_task **depends; /* list of tasks on which this task depends */
    int ndepends; /* length of dependency list */
    int nthreads; /* number of threads for this task */
    int numanode; /* preferred NUMA node for this task */
    ghost_task_flags flags; /* flags to configure this task */
} ghost_task;
\end{lstlisting}
The user-defined callback function containing the task's work and its arguments
have to be provided in the \texttt{func} and \texttt{arg} fields. The function's
return value will be stored by \ghost\ in \texttt{ret}. If the execution of this task
depends on the completion of \texttt{ndepends} other tasks, those can be specified as
a list of tasks called \texttt{depends}. The number of threads for the task has
to be specified in \texttt{nthreads}. Usually, a suitable number of PUs will be
reserved for this task. The field \texttt{numanode}
specifies the preferred NUMA node of PUs reserved for this task, which is
important for situations where different tasks work with the same data in main
memory in a process which spans several NUMA nodes. In this situation, and assuming a NUMA first touch policy, one can enforce a task which works on specific data to run
on the same NUMA node as the task which initialized this data.
The \texttt{flags} can be a combination of the following:
\begin{lstlisting}[style=MyCStyle]
typedef enum {
    GHOST_TASK_DEFAULT = 0; /* no special properties */
    GHOST_TASK_PRIO_HIGH = 1; /* enqueue task to head of the task queue */
    GHOST_TASK_NUMANODE_STRICT = 2; /* run task _only_ on the given NUMA node */
    GHOST_TASK_NOT_ALLOW_CHILD = 4; /* disallow child tasks to use task's PUs */
    GHOST_TASK_NOT_PIN = 8; /* neither reserve PUs nor pin threads */
} ghost_task_flags;
\end{lstlisting}

\piccaption{Program flow of an example application using a single \ghost\
    task for asynchronous task parallelism.\label{fig:ghost_tasking}}
\parpic[r]{\includegraphics[width=.45\textwidth]{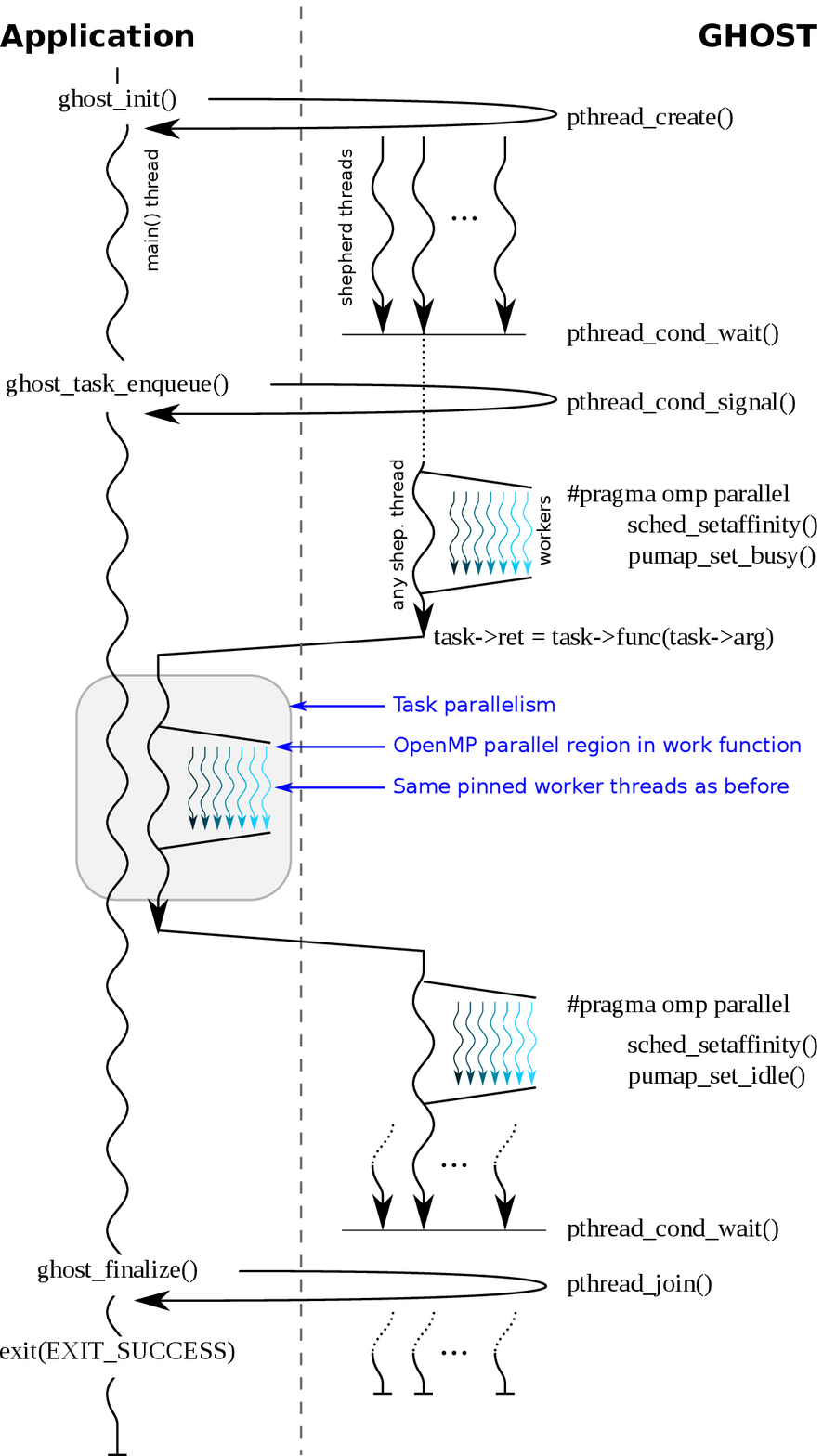}}
\Cref{fig:ghost_tasking} shows a simple flow chart of the execution of a \ghost\
application using task parallelism with a single task. 
In the
initialization phase \ghost\ creates a number of \emph{shepherd threads} which
will immediately wait on a condition. As a task gets enqueued,
this condition gets signalled
which causes an arbitrary shepherd thread to be woken up. 
Note that the \texttt{enqueue()}
function returns immediately. A decision whether this task can be executed
is now made by the shepherd thread based on the task's resource requirements. In
case the task can be executed, an
initial OpenMP parallel region is opened in which all threads of the task get
pinned to their exclusive PU and each PU is set to \emph{busy} in the
\texttt{pumap}. The task's work function now is called by the shepherd thread
and is executed in parallel to the user code which followed the call
to \texttt{enqueue()}. Due to physical persistence of OpenMP
threads, a parallel region in the task function will be executed by the same
threads as those which have been pinned by \ghost. 
Note that this persistence is not required by any standard. However, our
experiments have shown that the most relevant OpenMP implementations GOMP and
Intel OpenMP work like this, which makes the assumption of persistent OpenMP
threads realistic in practice.
Assuming that this assumption is invalid, one could still confine the children
of a shepherd thread to, e.g., a specific NUMA domain, by setting the shepherd
thread's affinity. This is possible since children
created via \texttt{fork()} inherit their parent's CPU affinity mask. Note that
this fallback mechanism involves a coarser level of affinity than the original
approach and might be inappropriate for scenarios where a task spans several
physical affinity (e.g., NUMA) domains.
After completion of the
task's work, the PUs are freed and threads are unpinned in another OpenMP parallel region.
At finalization time, the shepherd threads are terminated.

It is possible to create nested tasks, i.e., tasks running inside other tasks.
A parent task can be configured such that none of its children steals resources
of it by specifying the \texttt{GHOST\_TASK\_NOT\_ALLOW\_CHILD} flag. If
this flag is not set, it is expected that parents wait for their child tasks and
thus, the children can occupy the parent's resources (as demonstrated in the
task-mode SpMV example below).
In the simplest
case, there is only a single task which includes all the work done in the entire
application.
This ``main task'' should be created for all \ghost\ applications for controlled
thread placement and the avoidance of resource conflicts. Moreover, while
conducting
performance analyses using hardware performance counters, controlled placement
of threads is inevitable for making sense of the measurements.
On top of this, tasks can be used to implement task-level parallelism by
having several tasks running concurrently. Due to the fact that starting a task
is a non-blocking operation, asynchronous execution of work is inherently
supported by \ghost\ tasks. Normally, each task uses its own set of resources
(= PUs) which is not shared with other tasks. However, as mentioned above, a
task can also be requested to not reserve any compute resources. The PUs and their busy/idle state are
managed process-wide in a bitmap called \texttt{pumap}.
\new{The PUs available to \ghost\ can be set at the initialization
    phase. This feature can be used, e.g., for integration with third-party
resource managers that deliver a CPU set to be used.}

%\begin{wrapfigure}{L}{0.5\textwidth}
%\centering
%\includegraphics[width=.5\textwidth]{ghost_tasking.eps}
%\caption{Simple example of processing a single task in the application's main
%function.}
%\label{fig:ghost_tasking}
%\end{wrapfigure}
A realistic usage scenario for task level parallelism is communication hiding
via explicit overlap of communication and computation. This can be done,
e.g., in a parallel SpMV routine, which will be called task-mode SpMV. 
The following code
snippet shows the implementation of a task-mode SpMV using \ghost\ tasks.
\begin{lstlisting}[style=MyCStyle]
ghost_task *curTask, *localcompTask, *commTask;

/* get the task which I am currently in (will be split into child tasks) */
ghost_task_cur(&curTask);

/* create a heavy-weight task for computation of the local matrix part */
ghost_task_create(&localcompTask, &localcompFunc, &localcompArg, 
curTask->nthreads-1, GHOST_NUMANODE_ANY);

/* create a light-weight task for communication */
ghost_task_create(&commTask, &commFunc, &commArg, 1, GHOST_NUMANODE_ANY);

/* task-parallel execution of communication and local computation */
ghost_task_enqueue(commTask); ghost_task_enqueue(localcompTask);
ghost_task_wait(commTask); ghost_task_wait(localcompTask);

/* use the current (parent) task for remote computation */
remotecompFunc(remotecompArgs);

/* destroy the child tasks and proceed with current (parent) task */
ghost_task_destroy(localcompTask); ghost_task_destroy(commTask);
\end{lstlisting}
In this example, a main task is being split up into two child tasks.
Communication and local computation are being overlapped explicitly. In
principle, this could also be done via non-blocking MPI calls. However,
experience has shown that even nowadays some MPI
implementations do not fulfill non-blocking communication requests in an
asynchronous way. 
This has been discussed in various publications where also several attempts to
solving this problem have been proposed by, among others, Wittmann et
al.~\cite{Wittmann13} and Denis~\cite{Denis14}.
Thus, in order to create an assured overlap, independent of the
MPI library, \ghost's tasking mechanism could be used. Note that in many
application scenarios based on \ghost\ tasks, an MPI implementation supporting the
\texttt{MPI\_THREAD\_MULTIPLE} compatibility level is required.

\begin{figure}
\centering
\includegraphics[angle=-90,width=.8\textwidth]{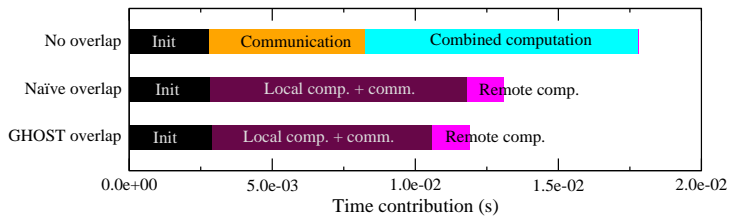}
\caption{Runtime contributions in different SpMV variants.
The ``No Overlap'' mode communicates input vector data
synchronously before it computes the full SpMV. The ``Overlap'' modes
communicate input vector data and at the same time compute the process-local
part of the SpMV. Here, the ``Na\"ive'' version relies on the asynchronicity of
non-blocking MPI calls, whereas the ``GHOST'' version uses explicit overlap by
means of \ghost\ tasks.}
\label{fig:spmv_taskbenefit}
\end{figure}
%\begin{wrapfigure}{R}{0.38\textwidth}
%\vspace{-10pt}
%\centering
%\includegraphics[width=.35\textwidth]{spmv_taskbenefit.eps}
%\caption{Contributions to solver runtime of different SpMV variants.}
%\label{fig:spmv_taskbenefit}
%\vspace{-20pt}
%\end{wrapfigure}
\Cref{fig:spmv_taskbenefit} depicts the potential performance gain by using
\ghost\ tasks. In this example, 100 parallel SpMV operations on 4 CPU-only compute nodes
using the
\texttt{vanHeukelum/cage15}\footnote{\url{http://www.cise.ufl.edu/research/sparse/matrices/vanHeukelum/cage15}} 
test case ($n = 5,154,859$, $n_\mathrm{nz} = 99,199,551$) stored in SELL-32-1024
have been performed. 
Note that both overlapped variants require a splitting of the
process-local matrix into a local and a remote part, where the remote part
contains entries with column indices who require communication of input vector
data.
Important observations are:
\begin{enumerate}[(i)]
    \item Overlapping communication and computation pays off in this case. The
        runtime for the two overlapped variants is significantly lower than for
        the non-overlapped variant. Note that this may not always be the case:
        The overlapped versions require the result vector to be stored twice and
        the
        cost of this may be higher than the benefit from communication hiding.
    \item The MPI library apparently features asynchronous point-to-point
        communication routines for this problem.
        The execution time for overlapped local computation and communication
        indicates that those operations are really overlapping. Note that this
        may not be the case in general, even for this MPI library. It is as well
        possible that the communication volume is below the ``eager limit'' and
        larger messages would not be transferred asynchronously.
    \item Affinity matters. Although one would not expect the task mode variant
        to perform any better than the MPI-overlapped variant, the execution
        time for local computation and communication is lower for the version
        using \ghost\ tasks. This can be explained by explicit thread placement.
\end{enumerate}

%\todo{explain different SpMV solver modes}

%% file: ghost_performance_features.tex
In this section we present several features of \ghost\ that constitute a
unique feature set leading to  high performance for a wide range of applications.
The goal of \ghost\ is neither to provide a ``Swiss army knife'' for sparse matrix
computations nor to re-invent the wheel. Instead, existing implementations are used 
and integrated into the \ghost\ interface whenever possible and feasible. 
In contrast to
the \emph{runtime} features presented in \cref{sec:runtime_features}, the
described performance features may be available in other libraries as well. In
order to justify their implementation in \ghost, short benchmarks or performance
models will be shown to demonstrate the potential or measurable benefit over
standard solutions.
%Note that no extensive performance study should be done in
%this section.

%% file: ghost_sparsemat_storage.tex
For the SpMV operation, the choice of a proper sparse matrix storage format is a crucial ingredient
for high performance.
In order to account for the heterogeneous design
of \ghost\ and simplify heterogeneous programming, \sellcs\ is chosen to
be the only storage format implemented in \ghost. This is no severe restriction,
since \sellcs\ can ``interpolate'' between several popular formats (see below).
We briefly review the format here.
A detailed and model-guided performance analysis of the SpMV kernel using the \sellcs\
format can be found in~\cite{Kreutzer14}. 

\sellcs\ features the hardware-specific tuning parameter C
and the matrix-specific tuning parameter $\sigma$. The sparse matrix is cut into
chunks, each containing C rows where C should be a multiple of the hardware's
SIMD width. In a heterogeneous environment, the relevant SIMD width should be
the maximum SIMD width over all architectures. For instance, considering our example node's
properties as given in \cref{tab:architectures} and using 4-byte values (single precision)
and indices, the minimum value of C should be
\vspace{-.5em}
$$[\max(32,64,128)\;\mathrm{Bytes}]/[4\;\mathrm{Bytes}] =
32.$$ 

\vspace{-.5em}
The rows in a chunk are padded with trailing
zeros to the length of the chunk's longest row. The chunk entries are stored
column-/diagonal-wise. Additionally, in order to avoid excessive storage
overhead for matrices with strongly varying row lengths,
$\sigma$ rows are sorted according to their nonzero count before chunk assembly.
\new{As this is a local operation, it can be trivially parallelized (which is also
done in \ghost).}
%\todo{\sellcs\ Bild?}
Note that, due to its general formulation, a range of further storage formats can be represented by \sellcs: 
\begin{itemize}
    \item SELL-1-1 \tabto{5em} $\widehat{=}$ \tabto{7em} CRS/CSR
    %\item SELL-$N$-$N$ \tabto{5em} $\widehat{=}$ \tabto{7em} JDS~\cite{Saad89}
    %\item SELL-C-$N$ \tabto{5em} $\widehat{=}$ \tabto{7em} pJDS~\cite{Kreutzer12}
    \item SELL-$n$-1 \tabto{5em} $\widehat{=}$ \tabto{7em} ITPACK/ELLPACK~\cite{Oppe87}
    \item SELL-C-1 \tabto{5em} $\widehat{=}$ \tabto{7em} Sliced ELLPACK~\cite{Monakov10}
\end{itemize}
%\todo{Ich dachte JDS ist spalten-weise und SELL zeilenweise? MK: "Nein,s.o."}
Thus, a considerable selection of known sparse matrix storage formats is supported by
\ghost. 
A single storage format for all architectures greatly facilitates
truly heterogeneous programming and enables quick (matrix) data
migration without conversion overhead.

\begin{figure}
\centering
\includegraphics[width=\textwidth]{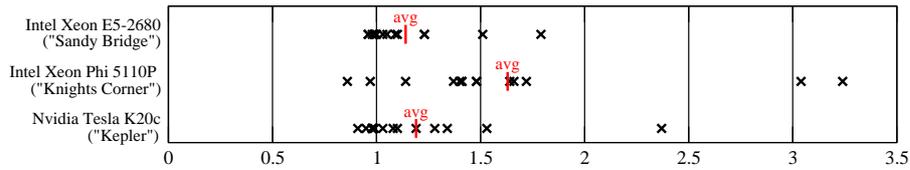}
\caption{Performance of SpMV using the unified \sellcs\ storage format relative to vendor-supplied
libraries with the device-specific data formats CRS for CPU/PHI and HYB for GPU
(figure taken from~\cite{Kreutzer14}).}
\label{fig:spmv_performance}
\end{figure}

\Cref{fig:spmv_performance} 
shows the relative performance of the \sellcs\
SpMV against the vendor-supplied libraries Intel MKL and NVIDIA cuSPARSE using
their device-specific sparse matrix
storage formats (CRS in MKL and HYB in cuSPARSE). 
It turns out that the performance of \sellcs\ is
on par with or better than the standard formats for most test matrices.

\new{For easy integration in existing software stacks, \ghost\ allows to construct a
\sellcs\ matrix from raw CRS data, i.e., row pointers, column indices, and
values. A common case in CS\&E applications is the subsequent appearance of multiple sparse
matrices with the same sparsity pattern but different values. 
Let us assume that we want to use \ghost\ and \sellcs\ for computations with
a CRS matrix obtained from another source.
Obviously, gathering
row lengths and column indices as well as the assembly of communication data structures and
permutation vectors only has to be done for the first read-in in this case.
Given the ML\_Geer matrix (cf.
\cref{sec:heterogeneous_execution}) present in CRS, we want to perform
SpMV using \ghost\ with SELL-32-128 on two CPU sockets with one MPI rank each.
We find that an initial complete construction of this matrix in \ghost\ (including
communication buffer setup and SELL permutation) costs as much as 48 SpMV
operations. Note that the communication buffer setup, which has to be done
independently of the library or the sparse matrix format, accounts for 78\% of the
construction time.
Each subsequent matrix construction only needs to update the matrix values.
Hence, all values need to be read from the CRS data and written to the \sellcs\
matrix. Taking into account the additional read operation due to the
write-allocate of the \sellcs\ matrix, we have at least $3\times n_\mathrm{nz}$ matrix
elements to transfer. The relative cost depends on the matrix data type.
Considering double precision data (and 32-bit indices), subsequent CRS-to-\sellcs\
conversions should cost as much as two SpMV operations. This performance can also
be observed in \ghost.
A possible future feature may be sparse matrix views. Using a view, a SELL-1-1
matrix could just point to existing CRS data and \ghost\ could be used for
computation with existing matrices at no conversion cost.
}

Note that \ghost\ differentiates between local and global indices. Even for
sparse system matrices of moderately large size, it may be necessary to have 64-bit integer numbers for
global indices. 
However, for the process-local part of the system matrix, 32-bit integers may still be
sufficient. As data movement should be minimized, especially for (often
bandwidth-bound) sparse solvers, it is possible to configure \ghost\ with
64-bit indices for global quantities (\texttt{ghost\_gidx}) and with 32-bit indices for local
quantities (\texttt{ghost\_lidx}).
Thus, the column and row indices of the entire process-local sparse matrix can be stored
using 32-bit integers.
Note that compression of the remote column indices as shown in
\cref{fig:sparsemat_procs} is inevitable in this case.
Considering the minimum amount of data transfers for the SpMV operation, using 32-bit instead of 64-bit
column indices for the sparse matrix results in a reduction of data transfers between 16 \% (complex double precision data) and 33 \%
(single precision data).

\new{It is also possible to incorporate matrix-free methods into \ghost. The SpMV
routine is stored as a function pointer in the \texttt{ghost\_sparsemat}. A
user can replace this function pointer by a custom function that performs the
SpMV in any (possibly matrix-free) way, while \ghost\ handles other kernels and communication of
vector data.}

%% file: ghost_blockvectors.tex
The architectural performance bottleneck for sparse linear algebra
computations is the main memory bandwidth for a wide range of algorithms.
Hence, reducing the movement of data through the memory interface often improves performance.
One well-known way to achieve this is to process
multiple vectors at once in a SpMV operation
if allowed by the algorithm.
%This technique has been successfully applied 
%decades.
Classic block algorithms are, e.g., the block Conjugate Gradient (CG)
method proposed by O'Leary et al.~\cite{O'Leary80} and the block GMRES
method introduced by Vital~\cite{Vital90}. The continued relevance of this
optimization technique is seen in recent publications, e.g.,
by R\"ohrig-Z\"ollner et al.~\cite{Roehrig14} in which the authors present
a block Jacobi-Davidson method.
Vector blocking is also very relevant in the field of eigenvalue solvers for
many inner eigenpairs.
For example, the FEAST algorithm~\cite{Polizzi09} and Chebyshev filter
diagonalization~\cite{Schofield12} profit from using block vector operations.
Basic work on potential performance benefits from using block vectors has been
conducted by Gropp et al.~\cite{Gropp99}, where a performance model for the Sparse
Matrix Multiple Vector Multiplication algorithm (SpMMV) has been established.
Support for block vectors (which are also represented by objects of
\texttt{ghost\_densemat}) has been implemented for many mathematical
operations in \ghost. 

Generally speaking, block vectors resemble tall and skinny dense matrices, i.e., matrices with
many rows and few columns. 
Although they are represented by general dense matrices, it has turned out that
existing BLAS implementations tend to deliver poor performance in
numerical kernels using tall and skinny dense matrices. This is the reason why
selected tall and skinny matrix kernels have been
implemented directly in \ghost.
%although general dense matrix (BLAS level 3) functions
%would also deliver correct results. 
Vectorized and fully unrolled versions of those kernels are automatically generated at compile time
for some predefined small dimensions. See \cref{sec:implementation} for
details on code generation and its impact on performance.

Let $V$ ($n \times m$) and $W$ ($n \times k$) be tall and skinny
dense matrices where $m,k \ll n$.
%\todo{bei Matrixgroessen wuerde ich besser
%kleine Buchstaben verwenden. MK: Andererseits ist die Matrixdimension ja auch
%``N'', und diese N hier ist ja das gleiche.} 
They are distributed in a
row-wise manner among the processes, similar to the system matrix in \cref{fig:sparsemat_procs}.
$X$ should be an $m \times k$ matrix which is redundantly stored on each
process. Three functions using tall and skinny dense
matrices are currently implemented in \ghost:
\begin{itemize}
    \item $X \gets \alpha V^T W + \beta X$ \\
        Tall skinny matrix transposed times tall skinny
        matrix (corresponds to inner product of block vectors): \texttt{ghost\_tsmttsm()}
    \item $W \gets \alpha VX + \beta W$ \\
        Tall skinny matrix times small matrix: \texttt{ghost\_tsmm()}
    \item $V \gets \alpha VX + \beta V$ \\
        In-place version of \texttt{ghost\_tsmm()}:
        \texttt{ghost\_tsmm\_inplace()}
\end{itemize}

\begin{figure}
    \centering
    \begin{minipage}[b]{.72\textwidth}
        \centering
        \subcaptionbox{$W \gets VX$}
        {\includegraphics[height=3cm,clip]{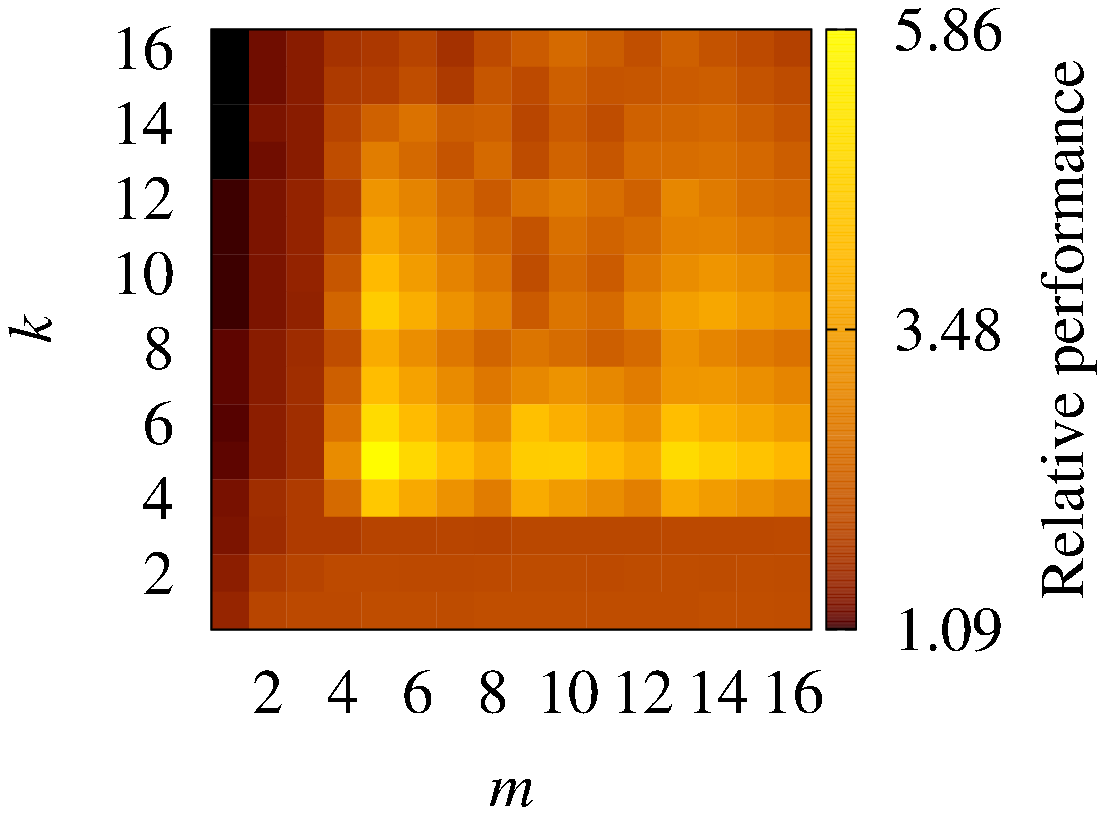}}
        \hfill
        \subcaptionbox{$X \gets V^TW$}
        {\includegraphics[height=3cm,clip]{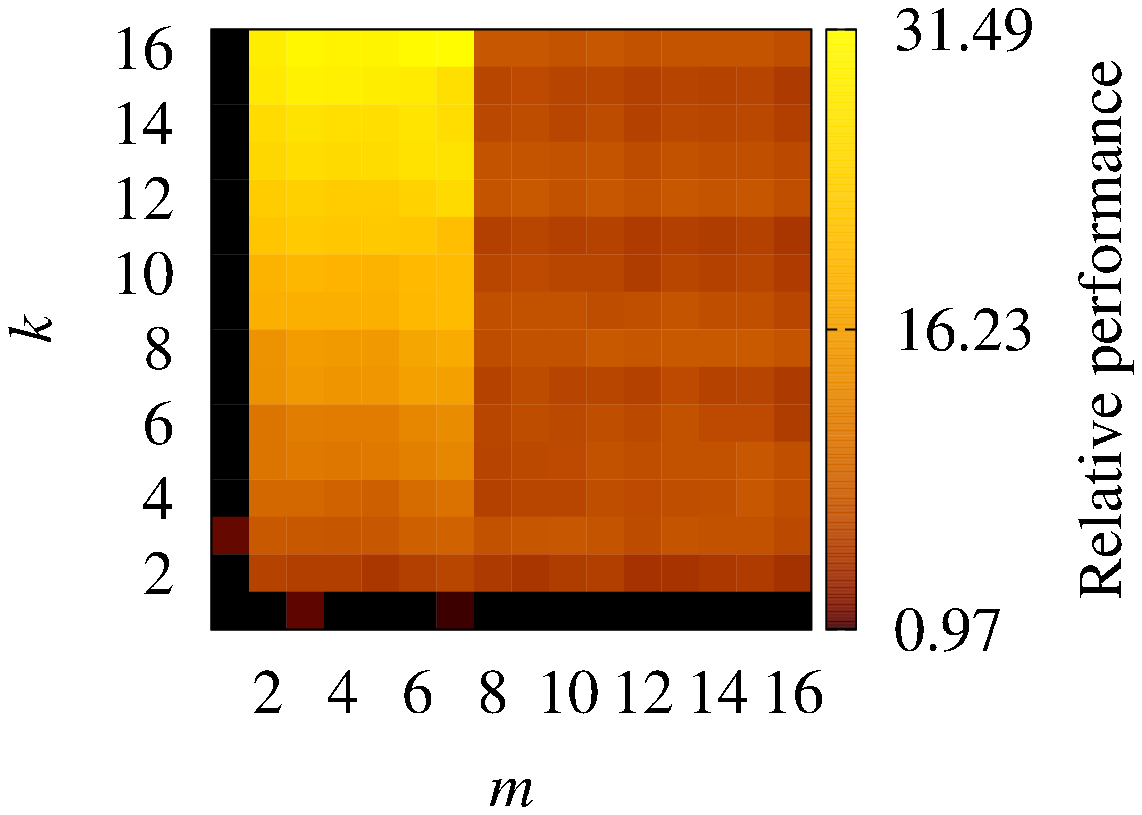}}
        \captionof{figure}{Speedup of custom tall and skinny matrix kernels over Intel MKL
        on a single CPU socket. $V$ is $n \times m$, $W$ is $n \times k$ and $X$
    is $m \times k$, where $m,k \ll n$.}
        \label{fig:special_gemm_performance}
    \end{minipage}%
    \hfill
    \begin{minipage}[b]{.23\textwidth}
        \centering
        \includegraphics[width=\textwidth,trim=0 -29.5mm 0 0,clip]{spmmv_rowcol.eps}
        \captionof{figure}{SpMMV performance of row- and col-major block
        vectors.}
        \label{fig:spmmv_rowcol}
    \end{minipage}
\end{figure}

%The specialized general dense matrix matrix multiplication (GEMM) kernels
%\texttt{ghost\_tsmm(\_inplace)()} and \texttt{ghost\_tsmttsm()} have only been
%implemented for the CPU and PHI until now.\comm{Solche Sachen muessen m.E. nicht unbedingt rein; diese Info aendert sich ja mit der Zeit} 
One may
assume that a mature library like the Intel MKL yields optimal
performance for a widely-used kernel like general dense matrix matrix
multiplication (GEMM) for matrices of any shape. However, this is not the case
as we demonstrate in \cref{fig:special_gemm_performance}. Similar observations
concerning performance drawback of MKL in the context of tall and skinny dense
matrices have
been made by Anderson et al.~\cite{Anderson11}. The GEMM kernel with
(not too small) square matrices can reach
a modern processor's peak floating point performance
if properly optimized.
The architectural performance bottleneck is the CPU's compute capability in this
case. However, this does not hold for tall and skinny matrices. The possibilities
of blocking are limited due to the small matrix dimensions in this case. This
results in a GEMM kernel which should ideally be memory-bound (as
long as the dimension $n$ is sufficiently large).
In \cref{fig:special_gemm_performance} it can be observed that the \ghost\
versions of both kernels are at least on par with the MKL performance for
relatively small dimensions. The potential speedup can be
up to 30$\times$ for some matrix
sizes.
Note that if the general function \texttt{ghost\_gemm()} is called, it 
first checks whether a suitable specialized function is applicable before
calling the BLAS library.

It is a known concern in extreme-scale computing that reduction operations are
susceptible to truncation errors.
In \ghost, the computation of the inner product of two tall skinny matrices
(\texttt{ghost\_tsmttsm()}) is one of the kernels where problems of this kind
may occur for very large $n$.
This motivated the addition of a variant of this kernel which uses Kahan
summation~\cite{Kahan65}. Depending on the width of the block vectors $m$ and
$k$, and hence the
computational intensity of the inner product, the overhead from the additional
floating point operations is small or negligible compared to
standard summation~\cite{Hofmann15}. However, the improvement in accuracy may be significant
which could result in a lower iteration count for some iterative algorithms and such a smaller time to solution.
This has been demonstrated, e.g., by Mizukami~\cite{Mizukami97} for CG-like methods.
%todo{Melven: habe ich Beispiele in phist, keine Spekulation. GH: Ja, aber dann muss da eine Referenz auf solche Arbeiten rein, die das zeigen.}
%It is even possible that accurate reduction is an essential ingredient for
%large-scale solver convergence.

In order to achieve a transparent support of block vectors in \ghost\ we implemented several BLAS level 1
routines and equipped them with  block vector support. Currently, the list of
block vector operations includes \texttt{axpy}, \texttt{axpby}, \texttt{scal}, and
\texttt{dot}. Each of those operations works vector-wise. In addition, versions
of \texttt{axpy}, \texttt{axpby}, and \texttt{scal} with varying scalar factors for each vector in the block have
been added: \texttt{vaxpy}, \texttt{vaxpby}, \texttt{vscal}.
Obviously, as block vectors are also represented as dense matrices, all of those
operations could be realized using existing BLAS level 3 routines. For example,
the \texttt{vscal} kernel could be replaced by a multiplication with a diagonal
matrix containing the scaling factors on its diagonal. However, this
would come at additional cost by transferring zeros, which we want to avoid.
Additionally, the concerns about the efficiency of BLAS level 3 operations for
tall and skinny dense matrices apply here as well.

\Cref{fig:spmmv_rowcol} shows benchmark results for the SpMMV kernel, comparing
row-major and column-major storage of block vectors with increasing width.
Storing the block vector in row-major corresponds to interleaved storage. As
expected, the performance for row-major storage surpasses the 
performance of column-major
storage due to a better data access pattern. This is well known, and both vendor-supplied sparse linear algebra libraries (Intel MKL and
NVIDIA cuSparse) support row-major block vectors in their SpMMV kernel.

%% file: ghost_kernelfusion.tex
Many sparse iterative solvers consist of a central SpMV routine
accompanied by several BLAS level 1/2 functions. It is thus useful
to
augment the SpMV with more operations according to our needs. The
general sparse matrix vector multiplication function
$y = \alpha (A - \gamma I) x + \beta y$ encompasses many of the practical
use cases. In \ghost\ this
operation can be chained with the computation of the dot products of
$\langle y,y \rangle$,
%\todo{Das SP sieht haesslich aus. Besser Macro machen mit Befehlen langle, rangle}
$\langle x,y \rangle$, and $\langle x,x \rangle$ as well as the BLAS level 1 operation $z = \delta z + \eta y$.
This approach is similar to the
well-known optimization technique of \emph{kernel fusion}.
Similar thoughts led to the addition of the so-called BLAS 1.5/2.5 operators
\texttt{AXPY\_DOT}, \texttt{GE\_SUM\_MV}, \texttt{GEMVT}, \texttt{TRMVT}, and
\texttt{GEMVER} to the updated BLAS standard~\cite{Blackford01}. 
%In some sense, \ghost's augmented SpM(M)V can be considered a Sparse BLAS 2.5 (3.5) routine.
%GH: Nicht plaudern
Siek et al.~\cite{Siek08} observed the
application specificity of these BLAS x.5 operators and made an attempt towards a
domain-specific compiler to generate arbitrarily fused kernels consisting of
different BLAS calls. This work has been continued by Nelson et
al.~\cite{Nelson15}, who plan to adapt their framework towards sparse matrices
in future work. 
Recently, the idea of kernel fusion has gained new attention in the GPU
programming community (\cite{Rupp14,Tabik14,Wahib14}).
%\todo{GH: War das nicht ein SC-Paper? Da sollte die Referenz jetzt verfuegbar sein}

The in- and output
vectors of the augmented SpMMV kernel are of the type general \texttt{ghost\_densemat}.
Hence, they may also be (views of)
block vectors. The values of $\alpha$, $\beta$, $\gamma$, $\delta$ and $\eta$ and the storage location of computed dot products are passed to the
function via a \texttt{ghost\_spmv\_opts} structure, which results in a single
interface function for any kind of (augmented) sparse matrix (multiple) vector 
multiplication.
Note that each augmentation on top of the standard SpM(M)V can be enabled
separately.
In the following we show a small example of how to use this function. 
\begin{lstlisting}[style=MyCStyle,mathescape]
/* default options for SpMV operation */
ghost_spmv_opts opts = GHOST_SPMV_OPTS_INITIALIZER;

/* compute $\color{lstcomment}y = Ax$ where $\color{lstcomment}y$ and $\color{lstcomment}x$ may be block vectors */
ghost_spmv(y,A,x,opts); 

/* compute $\color{lstcomment}y = (A - \gamma I) x$ with different $\color{lstcomment}\gamma$ for each vector in the block */
double shift[nvecs] = { /* init */ }; /* shift for each vector in the block */
opts.gamma = shift;
opts.flags |= GHOST_SPMV_VSHIFT;
ghost_spmv(y,A,x,opts);

/* compute $\color{lstcomment}y = (A - \gamma I) x - 2y$ with different $\color{lstcomment}\gamma$ for each vector, chained with $\color{lstcomment}x^Hy$ */
double dot[3*nvecs], beta = -2.0;
opts.beta = &beta;
opts.dot = dot;
opts.flags |= GHOST_SPMV_AXPBY|GHOST_SPMV_DOT_XY;
ghost_spmv(y,A,x,opts);
\end{lstlisting}
Both, the use of block vectors and kernel fusion, have large potential regarding performance, depending on the
algorithm. For example, for the Kernel Polynomial Method, a method for computing
the eigenvalue density of quantum systems as analyzed in
\cite{Kreutzer15}, a 2.5-fold performance gain for the overall solver could be
achieved by using block vectors and augmenting the SpMV.
Fused kernels are likely to be more cumbersome from an implementation point of
view than fine-grained kernels. For instance, fusing the SpMMV operation with block vector dot products
on a GPU leads to complex data access patterns which make an efficient
implementation hard to achieve (see~\cite{Kreutzer15} for details).
Due to the potentially high complexity of fused kernels and fundamental architectural differences,
hand-optimized implementations for each target architecture can hardly be avoided
if the focus is on high
efficiency in heterogeneous settings. 

A significant design decision for scientific computing libraries is whether and
how to use task and data parallelism. A task-parallel approach for
work distribution between heterogeneous devices, as implemented in
MAGMA~\cite{MAGMA}, may conflict with the presented optimization technique of
kernel fusion so that some optimization potential is left unused.
In cases where the potential benefits of task parallelism 
are limited, such as the sparse matrix algorithms targeted by \ghost,
data parallelism with kernel fusion may thus be favored over task parallelism.

%% file: ghost_implementation.tex
\begin{figure}
    \centering
    \begin{minipage}{.45\textwidth}
        \centering
        \includegraphics[width=\textwidth]{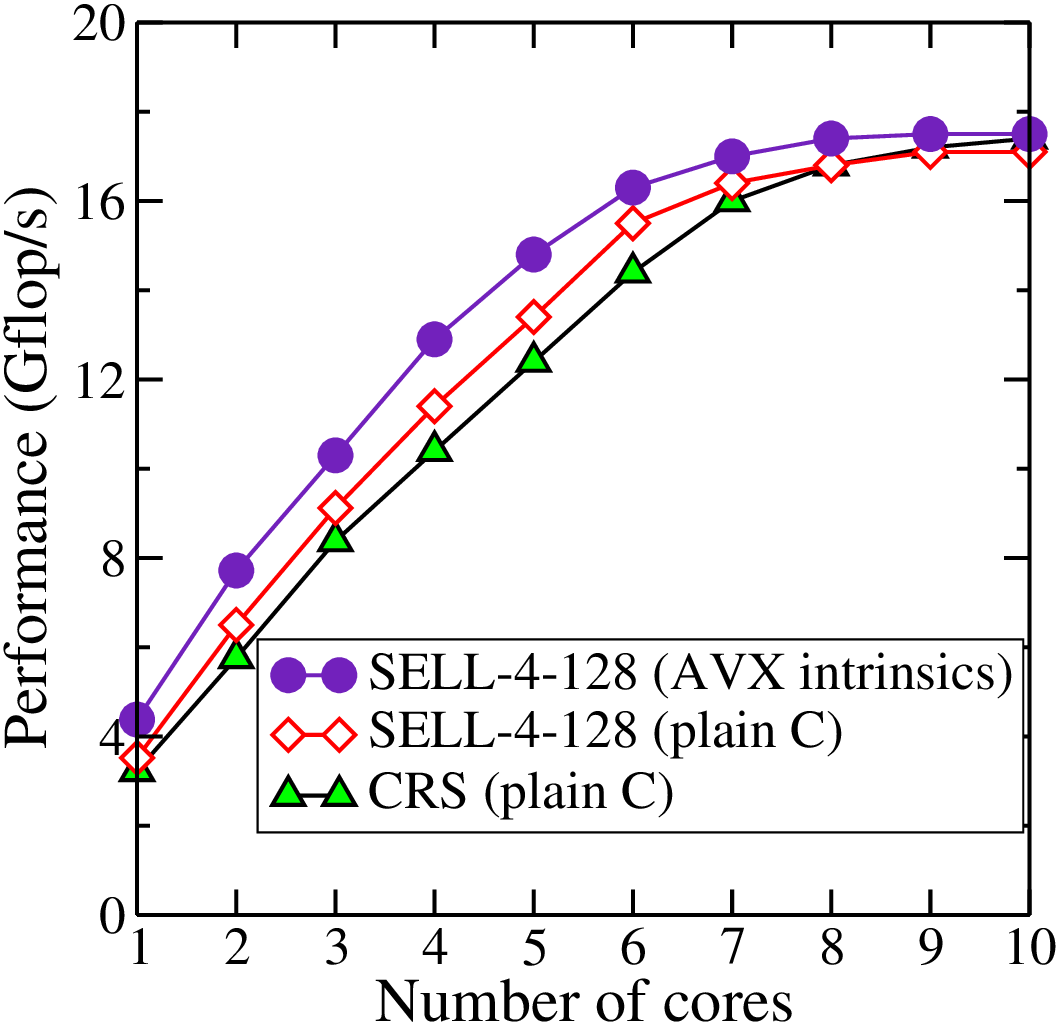}
        \captionof{figure}{Intra-socket performance on a single CPU showing the impact
        of vectorization on SpMV performance for different storage formats.}
        \label{fig:spmv_vectorization}
    \end{minipage}%
    \hfill
    \begin{minipage}{.45\textwidth}
        \centering
        \includegraphics[width=\textwidth]{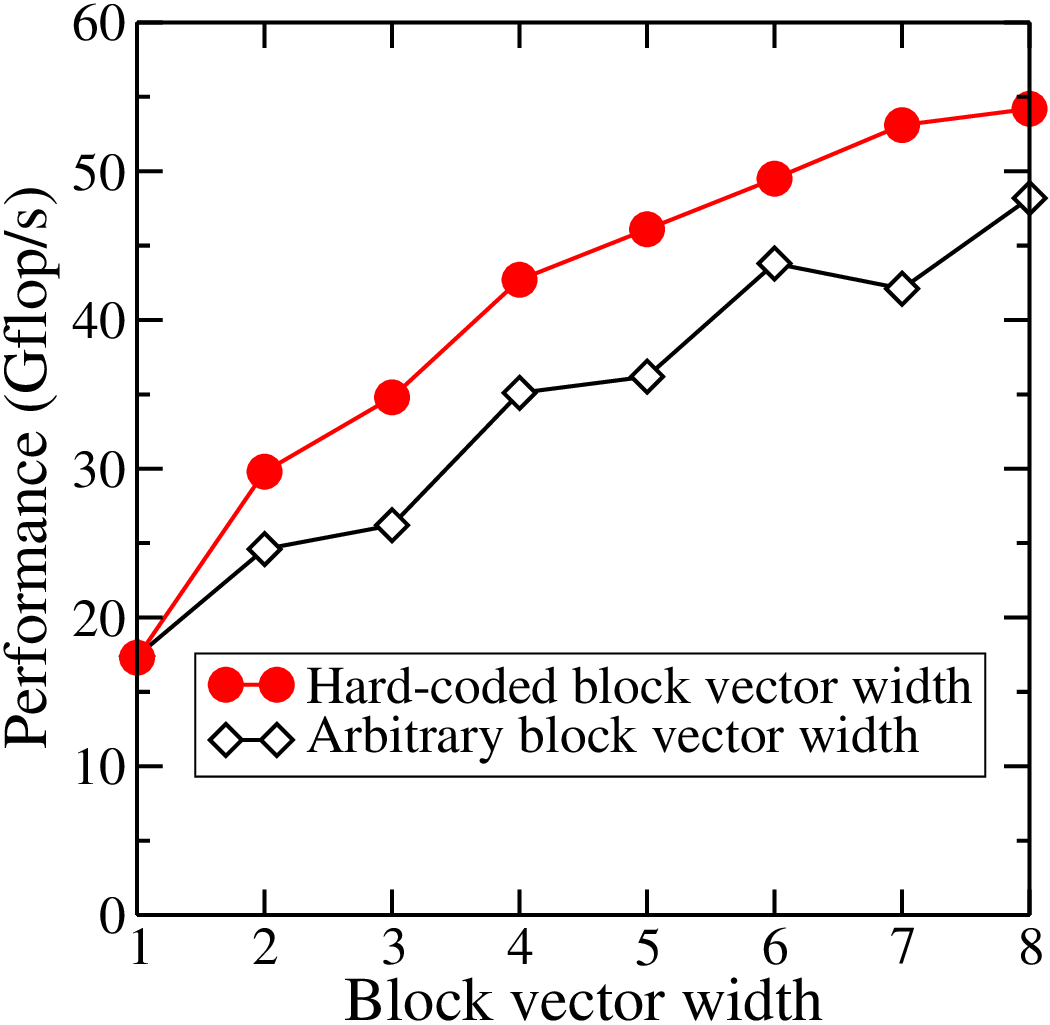}
        \captionof{figure}{The impact of hard-coded loop length on the SpMMV
    performance with increasing block vector width on a single CPU.}
        \label{fig:spmv_fixedloop}
    \end{minipage}
\end{figure}

\ghost{} is implemented with the goal of efficient execution from a
single core to the petaflop level. 
Modern CPUs feature SIMD 
units which cause code vectorization to be a
crucial ingredient for efficient core-level code. 
For kernels with sufficient
simplicity, automatic vectorization is likely to be done by the compiler. If
this is not the case, \ghost{} addresses this issue by using compiler pragmas, or SSE, AVX or MIC
compiler intrinsics for explicit vectorization.
Benchmarks on one CPU showing the impact of vectorization on SpMV performance can be seen in
\cref{fig:spmv_vectorization}. Here, we used the
\texttt{Sinclair/3Dspectralwave}\footnote{\url{http://www.cise.ufl.edu/research/sparse/matrices/Sinclair/3Dspectralwave.html}}
matrix ($n = 680,943$, $n_\mathrm{nz} = 30,290,827$) in complex double precision.
A first observation is that all three variants reach the same maximum performance
when using the full socket. Due to the the bandwidth-bound nature of the
SpMV kernel this limit corresponds to the CPU's maximum memory bandwidth.
However, the faster saturation of the explicitly vectorized SELL kernel allows
to use less cores to reach the same performance. The spare cores can be used
for working on independent tasks (cf. \cref{sec:resource_management}) or
they can be switched off in order to save energy. Hence, good vectorization
should always be a goal, even for bandwidth-bound kernels. Note that this is
especially true on accelerator hardware, where the width of vector units is
typically larger than on standard hardware (cf. \cref{tab:architectures}).

An obstacle towards efficient code
often observed by application developers
is lacking performance of existing program libraries due to their inherent and
indispensable requirement of generality. Often, better performance could be achieved if
performance-critical components were tailored to the application. Obviously,
this goal is opposing the original goal of program libraries, namely general
applicability.
An important feature of \ghost{} for achieving high performance is code
generation. At compile time, the user can specify common dimensions of data
structures for which
versions of highly-optimized kernels will be compiled. A prominent example is
the width of block vectors, i.e., the number of vectors in the block. This number
typically is rather small, potentially leading to overhead due to
short loops in numerical kernels.

The positive impact of hard-coded loop lengths on the performance of
SpMMV with increasing block vector width is demonstrated in
\cref{fig:spmv_fixedloop}. 
The hardware and problem setting is the same as
described above for \cref{fig:spmv_vectorization}, i.e., the performance for one
vector is the same as the saturated performance of \cref{fig:spmv_vectorization}.
If we configure the block vector widths $1,\dots,8$ at compile time a
significant performance benefit can be achieved compared to a variant where none of
them is configured. 
This is due to more optimization possibilities for the compiler due to simpler
code and a lower impact of loop overheads.
Note that for both variants the SELL chunk height C was
configured at compile time.

Code generation in \ghost{} serves two purposes. First, 
marked code lines can be duplicated with defined variations.
Second, it is possible to generate variations of functions, similar to C++ 
templates.
Assuming that we need a fully unrolled and AVX-vectorized kernel for scaling a
dense matrix with a scalar value. The template code using \ghost{} code
generation markers reads as follows:
\begin{lstlisting}[style=MyCStyle]
<@{\textbf{\#GHOST\_SUBST NVECS \${BLOCKDIM1}}}@>
void scale_unroll_avx_NVECS(ghost_densemat *a, double scal) {
  double *val = (double *)a->val;
  __m256d s = _mm256_set1_pd(scal);
  <@{\textbf{\#GHOST\_UNROLL\#}}@>__m256d tmp@ = _mm256_setzero_pd();<@{\textbf{\#(NVECS+3)/4}}@>
  for (ghost_lidx row=0; row < a->traits.nrows; row++) {
    <@{\textbf{\#GHOST\_UNROLL\#}}@>tmp@ = _mm256_load_pd(&val[row*a->stride+@*4]);<@{\textbf{\#(NVECS+3)/4}}@>
    <@{\textbf{\#GHOST\_UNROLL\#}}@>tmp@ = _mm256_mul_pd(tmp@,s);<@{\textbf{\#(NVECS+3)/4}}@>
    <@{\textbf{\#GHOST\_UNROLL\#}}@>_mm256_store_pd(&val[row*a->stride+@*4],tmp@);<@{\textbf{\#(NVECS+3)/4}}@>
  }
}
\end{lstlisting}
For simplicity, we assume appropriately padded and aligned input data and we
omit thread parallelization in this example.
Assuming that block vector widths 4 and 8 are specified in the build system,
the following two files are generated:\\
\begin{lstlisting}[style=MyCStyle]
void scale_unroll_avx_4(ghost_densemat *a, double scal) {
  double *val = (double *)a->val;
  __m256d s = _mm256_set1_pd(scal);
  __m256d tmp0 = _mm256_setzero_pd();
  for (ghost_lidx row=0; row < a->traits.nrows; row++) {
    tmp0 = _mm256_load_pd(&val[row*a->stride+0*4]);
    tmp0 = _mm256_mul_pd(tmp0,s);
    _mm256_store_pd(&val[row*a->stride+0*4],tmp0);
  }
}
\end{lstlisting}
\begin{lstlisting}[style=MyCStyle]
void scale_unroll_avx_8(ghost_densemat *a, double scal) {
  double *val = (double *)a->val;
  __m256d s = _mm256_set1_pd(scal);
  __m256d tmp0 = _mm256_setzero_pd();
  __m256d tmp1 = _mm256_setzero_pd();
  for (ghost_lidx row=0; row < in->traits.nrows; row++) {
    tmp0 = _mm256_load_pd(&val[row*a->stride+0*4]);
    tmp1 = _mm256_load_pd(&val[row*a->stride+1*4]);
    tmp0 = _mm256_mul_pd(tmp0,s);
    tmp1 = _mm256_mul_pd(tmp1,s);
    _mm256_store_pd(&val[row*a->stride+0*4],tmp0);
    _mm256_store_pd(&val[row*a->stride+1*4],tmp1);
  }
}
\end{lstlisting}
Note that in this example the factor for code duplication \texttt{(NVECS+3)/4} depends on the
function variant. This disqualifies the straightforward use of C++ templates
for function variation, as values of the template parameters would have to be
known before compilable (i.e., with evaluated \texttt{GHOST\_UNROLL} statements)
code is present. While this might still be possible with more sophisticated C++
techniques, it cannot be denied that the possibility to look at the high-level
intermediate representation as shown in the example above is an advantage of our
approach over C++ metaprogramming.
Note that it is not always possible to replace the generation of code line
variants by loops, e.g., for the declaration of variables.

Fallback implementations exist for all compute kernels. This guarantees general
applicability of \ghost\ functions. The degree of
specialization gets diminished until a suitable implementation is found, which
probably implies a performance loss. For
example, if a kernel is not implemented using vectorization intrinsics and hard-coded
small loop dimensions, a version with one arbitrary loop dimension is searched. 
If none of the small loop dimensions is available in an
explicitly vectorized kernel, \ghost{} checks for the existence of a high-level language implementation with hard-coded
small loop dimensions, and so on. In case no
specialized version has been built into \ghost{} the library will select the
highly-general
fallback version.

%% file: ghost_integration.tex
%\todo{Discuss matrix-free methods in this section}
In this section we want to briefly discuss how \ghost\ can be used with existing sparse 
solver libraries. A characteristic feature of typical iterative solvers for sparse linear 
systems or eigenvalue problems is that they require only the application of the matrix to a given vector.
It is therefore good practice to separate the implementation of such methods from the data 
structure and details of the SpMV.

One approach that originated in the days of Fortran 77 is the
`Reverse Communication Interface' (RCI). The control flow passes back
and forth between the solver routine and the calling program unit, which
receives instructions about which operations are to be performed on which data in memory.
While this programming model is still widely used in, e.g., ARPACK~\cite{lehoucq98arpack}
and even in modern libraries such as Intel MKL~\cite{MKL}, it is awkward and error-prone
by today's standards. Another idea is to use callback functions for selected operations. For 
example, the eigensolver package PRIMME~\cite{PRIMME} requires the user to provide the 
SpMMVM and a reduction operation on given data. 

Neither RCI nor simple callbacks can make optimal use of \ghost. 
Obviously such software could only make use of accelerators by means of
offloading inside a function scope. If no special attention is paid to data
placement, this is typically inefficient due to the slow PCI express bus between CPU and device.
%\todo{Wenn man das richtig implementiert koennte man sich die Transfers ja auch
%sparen...}
Even on the CPU, 
\ghost\ preferably would control memory allocation itself to achieve alignment and NUMA-aware 
placement of data. Another drawback is the restriction to data structures as
prescribed by such solvers. For instance, the required storage order of block
vectors is 
typically column-major, which may be also inefficient (cf.
\cref{sec:blockvectors}).

The Trilinos package Anasazi~\cite{baker2009anasazi} takes a different approach. It requires 
the user to implement what we call a `kernel interface', an extended set of callback 
functions that are the only way the solver can work with matrices and vectors. New (block) 
vectors are created by cloning an existing one via the kernel interface.
Thus, memory allocation stays on the user side and can be done, e.g., on a GPU, with a 
custom data layout, or applying further optimizations. 

In the iterative solver package
PHIST~\cite{PHIST} we use a 
similar kernel interface which is written in plain C. It does not 
require a very general vector view concept and has some functionality for executing (parts 
of) kernels asynchronously as \ghost\ tasks. PHIST also provides \ghost\ adapters for Anasazi 
and the linear solver library Belos from Trilinos, with the restriction that 
views of permuted columns of a block vector do not work with row-major storage. This is 
not a grave restriction as the feature is -- to our knowledge -- hardly used in the packages.
For a performance study of the block Jacobi-Davidson method implemented in PHIST 
(using \ghost), see~\cite{Roehrig14}.

%% file: ghost_case_study.tex
\new{We have demonstrated the applicability and performance of \ghost\ in a number of
publications. In~\cite{Roehrig14}, we have presented and implemented a block
Jacobi-Davidson method using PHIST \& \ghost\
on up to 512 dual-socket CPU nodes. A fully heterogeneous
\ghost\ implementation of the kernel polynomial method which we scaled up to
1024 CPU+GPU nodes has been demonstrated in~\cite{Kreutzer15}. In the meantime, we have
continued our scaling studies of this application to 4096 heterogeneous nodes.
Recent work includes the implementation of Chebyshev filter diagonalization, for
which we show performance data on up to 512 dual-socket CPU nodes in~\cite{Pieper15}.

While all of the presented work has been conducted within activities closely
related to
the \ghost\ project, we see that it is of special interest for a broader
potential user base
how \ghost\ could integrate in existing CS\&E software stacks.
In the following we want to demonstrate the applicability and performance of
\ghost\ using
the Krylov-Schur method~\cite{Stewart02} for finding a few eigenvalues of large sparse matrices.
An implementation of this method is available in the Anasazi package~\cite{baker2009anasazi} of Trilinos.
As mentioned in the previous section, PHIST can serve as an interface layer
between algorithmic packages like Anasazi and kernel libraries like
\ghost\ or Tpetra (+Kokkos). Developers can thus work at a high level of
abstraction and have the option to switch between kernel implementations.
For this study, we are using version 11.12.1 of Trilinos and
an MPI+X approach with OpenMP parallelization on the socket level.
The test case is the non-symmetric
MATPDE\footnote{\url{http://math.nist.gov/MatrixMarket/data/NEP/matpde/matpde.html}}
problem. It represents a five-point central finite difference discretization
of a two-dimensional variable-coefficient linear elliptic equation on an 
$n \times n$ grid with Dirichlet boundary conditions. The ten eigenvalues with largest
real part are sought using a search space of twenty vectors. The convergence
criterion is a residual tolerance of $10^{-6}$. We set the random number
seed in \ghost\ in a way which guarantees consistent iteration counts between successive
runs.

\begin{figure}
    \centering
    \begin{subfigure}[t]{0.45\textwidth}
        \includegraphics[width=\textwidth]{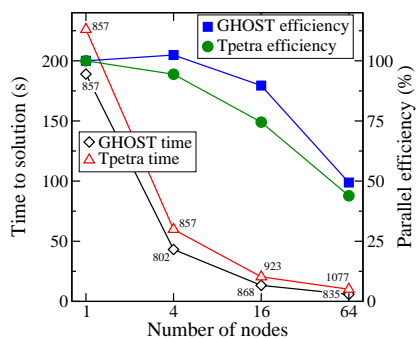}
        \caption{Strong scaling runtime (left axis) and parallel efficiency
            (right axis) for $n=2^{12}$.}
            \label{fig:ks_strongscaling}
    \end{subfigure}
    \hfill
    \begin{subfigure}[t]{0.45\textwidth}
        \includegraphics[width=\textwidth]{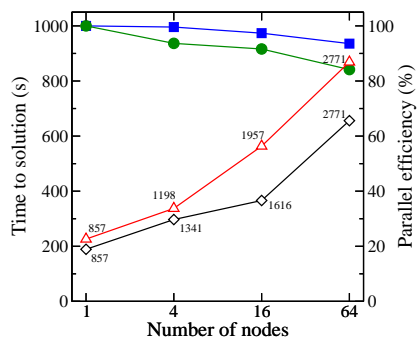}
        \caption{Weak scaling runtime (left axis) and parallel efficiency (right
            axis) for $n=2^{12,\dots,15}$.}
            \label{fig:ks_weakscaling}
    \end{subfigure}
    \caption{Scaling behavior of GHOST and Tpetra on up to 64 dual-socket CPU
    nodes for Anasazi's implementation of the Krylov-Schur method. The
annotations show the number of iterations until convergence. 
The computed parallel efficiencies consider changed iteration counts.}
    \label{fig:ks}
\end{figure}

GHOST integrates well with Anasazi and is straightforward to use on this level.
Moreover we show in 
\cref{fig:ks} that \ghost\ surpasses Tpetra both in
terms of performance and scalability. On a single node one can save about 16\%
of runtime for the entire solver. \Cref{fig:ks_strongscaling} reveals a
higher parallel efficiency of \ghost. Consequently, the better node-level
performance gets amplified on larger node counts, resulting in a 42\% runtime
saving on 64 nodes. For weak scaling, similar conclusions can be drawn from
\cref{fig:ks_weakscaling}. At the largest node count, the parallel efficiency
of \ghost\ is ten percentage points above Tpetra.
Relevant \ghost\ features used in the presented runs are resource
management (thread pinning), SpMV with \sellcs\ and auto-generated kernels for tall \& skinny
dense matrix multiplications.
Note that even higher performance could possibly be obtained by exploiting
advanced algorithmic optimizations available with \ghost, such as kernel fusion,
block operations and communication hiding. However, those would potentially
require a re-formulation of the algorithm which is not what we wanted to
demonstrate here.
}

%\begin{table}[tb]
%\centering
%\begin{tabular}{l|cccc}
%\hline
%Kernel library & \# CPU sockets & Execution mode & Iterations & t\_tot \\
%\hline
%\hline
%Tpetra & 2xCPU & 1055 & 134.2 \\
%GHOST & 2xCPU & 1231 & 110.5 \\
%\hline
%\end{tabular}
%\caption{Iteration count and runtime for solving the Krylov-Schur method for a
%$2048 \times 2048$ MATPDE test case. The ten eigenvalues with largest
%real part are searched using a search space of twenty vectors. The convergence
%criterion is a residual tolerance of 1E-6.}
%\label{tab:case_study_times}
%\end{table}

%% file: ghost_conclusions.tex
\ghost\ is a novel and promising attempt towards highly scalable heterogeneous sparse
linear algebra software. 
It should not be considered a comprehensive library but rather a toolbox 
featuring approaches to the solution of several problems which we have identified as
relevant on modern hardware in the context of sparse solvers.
A crucial component of highly efficient software, especially in the complex
environment of heterogeneous systems, is sensible resource management.
Our flexible, transparent, process-based and data-parallel approach for heterogeneous execution is
accompanied by a lightweight and affinity-aware tasking mechanism, which
reflects the requirements posed by modern algorithms and hardware architectures.
During the ongoing development, we have observed that high performance is the
result of a mixture of ingredients. First, algorithmic choices and optimizations
have to be made considering the relevant hardware bottlenecks. In the context
of sparse solvers, where minimizing data movement is often the key to higher
efficiency, this includes, e.g., vector blocking and kernel fusion.
Second, while implementing those algorithms, it is crucial to have an idea of upper performance bounds.
This can be accomplished by means of performance models, which form a
substantial element of our development process. This is demonstrated
in~\cite{Kreutzer14} and~\cite{Kreutzer15}.
An optimal implementation may come at the cost of fundamental
changes to data structures, e.g., storing dense matrices as row- instead of
column-major or changing the sparse matrix storage format from CRS to \sellcs. 
During the ongoing development it has turned out that often the
generality of the interface has to be traded in for high performance.
There are several ways to relax this well-known dilemma. Very promising is,
e.g., a close
collaboration between library and application developers with the possibility
for the latter to feed their application-specific knowledge into the library.
In \ghost\, this idea is implemented by automatic code generation.

%% file: ghost_outlook.tex
In its current state, \ghost\ has no provision for exploiting matrix symmetry.
Obviously, there is large potential for increased of performance if symmetric (or
Hermitian) matrices were treated as such. The implementation is
challenging, but cannot be avoided in the long run. Bringing sparse
solvers to a very large scale is often limited by malicious sparse matrix
patterns which lead to communication dominating the runtime. This can be
ameliorated by bandwidth reduction of the sparse matrix. A goal for further
development is the evaluation and implementation of additional bandwidth reduction
algorithms like, e.g., hypergraph partitioning~\cite{Devine06}. Furthermore, the
optimization of heterogeneous MPI communication, e.g., using GPUdirect which
bypasses the host memory in GPU-GPU communication, should be investigated in order
to improve the communication performance. 
Future architectural developments, like deeper memory hierarchies and a tighter
integration of ``standard'' and ``accelerated'' resources require rethinking
existing performance models and possibly new implementations. 
Currently, the heterogeneous work distribution weights have to be specified by the user,
mostly based on knowledge about the involved hardware architectures and their
capabilities. In future work, micro-benchmarks will be integrated into \ghost\
that allow automatic determination of device-specific work weights.
On top of that, another important goal for future development is dynamic and automatic
load balancing during an iterative solver's runtime. Currently, the sparse matrix
portion for each process is fixed during the entire runtime.
By using the \sellcs\ storage format, it will be straightforward to
communicate matrix data at runtime between heterogeneous devices to overcome
load imbalances.